\documentclass[letterpaper,twocolumn,10pt]{article}

\usepackage[table,dvipsnames]{xcolor}
\usepackage{usenix}
\usepackage{amsmath,amssymb,amsthm,bm}
\usepackage{graphicx}
\usepackage{subcaption}
\usepackage{booktabs,multirow,makecell,tabularx,array,float,placeins}
\usepackage{pifont}

\theoremstyle{remark}

\definecolor{cgreen}{rgb}{0,0.6,0}
\newcolumntype{Y}{>{\raggedright\arraybackslash}X}
\newcommand{\cmark}{\textcolor{cgreen}{\ding{52}}}
\newcommand{\xmark}{\textcolor{red}{\ding{55}}}

\newcommand{\bx}{\bm{x}}
\newcommand{\bp}{\bm{p}}
\newcommand{\bg}{\bm{g}}
\newcommand{\bdelta}{\bm{\delta}}
\newcommand{\bmu}{\bm{\mu}}
\newcommand{\bSigma}{\bm{\Sigma}}
\newcommand{\bQ}{\bm{Q}}

\title{Checkerboard: Closed-Form and Data-Independent Trigger Design for Clean-Label Backdoor Attacks}
\author{
	\rmfamily
	Yi Yang$^{1}$ \quad
	Jinyang Huang$^{1}$ \quad
	Binbin Liu$^{1}$ \quad
	Feng-Qi Cui$^{2}$ \quad
	Xiaokang Zhou$^{3}$
	\\
	Haiming Jin$^{4}$ \quad
	Zhi Liu$^{5}$ \quad
	Jie Zhang$^{6*}$ \quad
	Meng Li$^{1*}$
	\\
	$^{1}$Hefei University of Technology
	\quad
	$^{2}$University of Science and Technology of China
	\\
	$^{3}$Kansai University
	\quad
	$^{4}$Shanghai Jiao Tong University
	\\
	$^{5}$The University of Electro-Communications
	\quad
	$^{6}$Agency for Science, Technology and Research (A*STAR)
}
\date{}

\begin{document}
\maketitle

\begingroup
\renewcommand{\thefootnote}{\fnsymbol{footnote}}
\footnotetext[1]{Corresponding authors.}
\endgroup

\begin{abstract}
Backdoor attacks threaten the deep-learning supply chain by poisoning a small fraction of the training data so that a model behaves normally on clean inputs but maps triggered inputs to an attacker-chosen class. Clean-label backdoor attacks are especially difficult to audit because poisoned examples preserve their semantic labels. Yet existing clean-label attacks often require surrogate model training, auxiliary data access, or iterative optimization.

In this paper, we present \emph{Checkerboard}, a clean-label backdoor attack with closed-form, data-independent trigger design. We formulate trigger design through an input-space Fisher-separability objective and, under a ridge four-neighbor local-smoothness prior for natural images, obtain the pixel-wise checkerboard as a closed-form maximizer of the resulting design proxy without data access, model training, or optimization. Across four benchmark datasets, Checkerboard outperforms the evaluated norm-bounded clean-label attacks and achieves state-of-the-art performance under low global poisoning rates. For example, on CIFAR-10, under a trigger perturbation of $10/255$, poisoning 20 training samples achieves $95.72\%$ Attack Success Rate (ASR). On IN-100, a global poisoning rate of only $0.4\%$ yields over $83\%$ ASR without degrading clean accuracy. The proposed attack also remains effective against state-of-the-art backdoor defenses and shows resistance to adaptive defenses under simple modification.

\end{abstract}

\section{Introduction}
\label{sec:introduction}

Backdoor attacks pose a severe threat to the machine-learning supply chain~\cite{li2022backdoor}. By poisoning a small fraction of the training data, an adversary can cause a model to behave normally on clean inputs while misclassifying inputs containing a specific trigger to an attacker-chosen target class. Because the manipulation is dormant until the trigger appears, backdoor attacks are particularly difficult to diagnose. Consequently, backdoor attacks are considered one of the most severe threats against deep-learning systems~\cite{baracaldo2022machine,goldblum2022dataset,tao2024exploring,vassilev2024adversarial}.

Classic backdoor attacks are typically \textbf{dirty-label}: the attacker injects a trigger into selected samples and maliciously relabels them to the target class~\cite{gu2017badnets,chen2017targeted,qiu2024belt,ma2024watch,cheng2024lotus,Liu2024LADDERMB,xu2025towards,lin2025revisiting}. While highly effective, such attacks are less stealthy because the poisoned samples exhibit explicit label inconsistency and are therefore more vulnerable to human inspection and data filtering. In contrast, \textbf{clean-label} backdoor attacks preserve the original labels of poisoned samples, maintaining semantic consistency between image content and annotation~\cite{turner2019label}. This label consistency makes clean-label poisoning harder to audit.

\textbf{Existing work on clean-label attacks:} Label consistency comes at a cost. The trigger must compete with benign target-class semantics during training because poisoned samples retain their original labels. To make the trigger learnable, many effective clean-label attacks rely on attack-specific learning or optimization. LC, HTBA, and SAA construct poisons through adversarial, feature-collision, or gradient-matching objectives~\cite{turner2019label,saha2020hidden,souri2022sleeper}. Narcissus, GenBound, Invisible Poison, COMBAT, Hybrid, and BAAT use surrogate models, learned generators, auxiliary information, joint optimization, or semantic editing in different combinations~\cite{zeng2023narcissus,yu2024generalization,ning2021invisible,huynh2024combat,yuan2025stealthy,zhu2025towards}. These approaches demonstrate the strength of optimized clean-label poisoning, but their construction can involve substantial computation, data access, and design complexity.

Learning-free poisoning applies a fixed trigger to target-class samples without requiring surrogate training or iterative optimization to derive the trigger pattern. This is attractive because it is simple and efficient. However, under the clean-label constraint, existing learning-free approaches \cite{barni2019new} are much less effective at low poisoning budgets. Their trigger patterns are typically heuristic rather than analytical, and therefore fail to induce a sufficiently strong and consistent feature shift for the model to learn. This leaves an important question: \textbf{can a learning-free clean-label backdoor attack remain highly effective under low poisoning budgets?}

\textbf{Challenges:} Answering this question requires overcoming two challenges. First, the trigger must be constructed without training a surrogate model, ruling out victim or surrogate access, attack-side training, iterative optimization, and empirical data access. Second, under the clean-label constraint and a bounded target-class poisoning budget, the fixed trigger must induce a sufficiently distinctive and consistent feature shift for the victim to learn despite the dominance of benign target-class semantics. The attack must therefore combine data-independent trigger design with empirical learnability from an attacker-controlled subset of target-class samples.

\begin{figure}[t]
  \centering
  \includegraphics[width=\columnwidth]{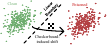}
  \caption{Trigger construction from the input-space separability view. Under the stated natural-image graph prior, maximizing the adopted proxy yields the pixel-wise checkerboard.}
  \label{fig:separability-intuition}
\end{figure}

\textbf{Approach Overview:} We answer this question with \textbf{Checkerboard}, a clean-label and learning-free backdoor attack with a closed-form, data-independent trigger design. As illustrated in Figure~\ref{fig:separability-intuition}, our starting point is a distribution-separability view of trigger design: a bounded fixed pattern should induce a consistent shift that is large relative to natural-image variation. We instantiate this idea through an input-space Fisher discriminant-ratio (FDR) proxy between a reference clean-image distribution and the triggered distribution obtained by applying a candidate pattern. Because the natural-image precision is unavailable to the attacker, we replace it with a generic ridge plus a weighted four-neighbor graph-Laplacian prior. Under further luminance restriction, local-smoothness model, and four-neighbor image grid, the resulting objective is globally maximized in closed form by a pixel-wise checkerboard pattern, up to sign. This analysis identifies the checkerboard as the closed-form optimum under the adopted input-space proxy and prior. As a result, the checkerboard trigger is derived analytically rather than optimized, its construction requires no surrogate model, attack-specific training, numerical optimization or data access.

\textbf{Evaluation Overview:} We evaluate Checkerboard on CIFAR-10, CIFAR-100, CelebA-8, and IN-100. Our results show that even under a constrained clean-label threat model, a learning-free attacker can successfully implant highly effective backdoors. Checkerboard consistently outperforms prior norm-bounded clean-label attacks. For example, with a perturbation budget of \(10/255\), poisoning only 20 samples (\(0.04\%\) global poisoning rate) on CIFAR-10 yields up to \(95.72\%\) Attack Success Rate (ASR), with a perturbation budget of \(16/255\) poisoning 520 samples (\(0.4\%\) global poisoning rate) on IN-100 yields over \(83\%\) ASR. These results suggest that strong clean-label backdoors do not necessarily require surrogate-model training or expensive trigger optimization, substantially reducing the attack-side knowledge and computation required for clean-label poisoning.

Low poisoning rates also reduce the fraction of modified training data. Together with the low train-time perturbation, it keeps the resulting poisons visually consistent with clean data. We evaluate the attack against 7 representative defenses (with an additional 12 defenses reported in the appendix) from model- and data-level families, and find that no defense consistently removes the backdoor without incurring substantial utility loss. We further show that Checkerboard remains effective under adaptive defenses with simple modification.

In summary, this paper makes the following contributions:
\begin{itemize}
  \item We introduce Checkerboard, a simple, norm-bounded, clean-label backdoor attack with a fixed, data-independent trigger design. Trigger construction requires no surrogate model training, auxiliary data access or numeric optimization.

  \item We formulate fixed-trigger design using an input-space FDR proxy and show that, under the luminance restriction and a ridge-regularized, positively weighted four-neighbor graph-Laplacian prior, the pixel-wise checkerboard is a closed-form global maximizer, up to sign.

  \item We show that Checkerboard achieves strong attack success across 4 benchmarking datasets, while remaining difficult to mitigate with a broad range of defenses.
\end{itemize}

\section{Background and Related Work}
\label{sec:related}

Backdoor attacks compromise a victim model by injecting a small number of poisoned samples into the training data. Consequently, the trained model behaves normally on clean inputs but maps triggered inputs to an attacker-chosen target class~\cite{li2022backdoor}. A standard distinction in the literature is between \textit{dirty-label} and \textit{clean-label} attacks, depending on whether poisoned samples remain consistent with their labels.

\subsection{Dirty-Label Attacks}

Most classic backdoor attacks are dirty-label. A common strategy is to stamp selected samples with a trigger and relabel them to the target class~\cite{li2022backdoor,zhang2024exploring}. Because the poisoned samples are explicitly mislabeled, the trigger often becomes a strong discriminative shortcut during training. However, such attacks are generally less stealthy, since the semantic content of the poisoned image no longer matches its assigned label and may raise suspicion during inspection~\cite{turner2019label}.
\subsection{Clean-Label Attacks}

Clean-label attacks preserve the original labels of poisoned samples. This makes the attack more challenging because the injected trigger must compete with benign target-class semantics during training. We organize prior clean-label attacks into two categories: \textit{learning-based} and \textit{learning-free} methods.

\begin{table}[t]
	\centering
	\caption{Compared properties of clean-label attacks. }
	\label{tab:related-properties}
	\small
	\setlength{\tabcolsep}{3.5pt}
	\begin{tabular}{lccc}
		\toprule
		Attack & Learning-free & Data-free & Closed-form \\
		\midrule
		LC~\cite{turner2019label}                    & \xmark & \xmark & \xmark \\
		Invisible Poison~\cite{ning2021invisible}   & \xmark & \xmark & \xmark \\
		HTBA~\cite{saha2020hidden}                  & \xmark & \xmark & \xmark \\
		SAA~\cite{souri2022sleeper}                 & \xmark & \xmark & \xmark \\
		Narcissus~\cite{zeng2023narcissus}          & \xmark & \xmark & \xmark \\
		GenBound~\cite{yu2024generalization}        & \xmark & \xmark & \xmark \\
		Hybrid~\cite{yuan2025stealthy}              & \xmark & \xmark & \xmark \\
		COMBAT~\cite{huynh2024combat}               & \xmark & \xmark & \xmark \\
		BAAT~\cite{zhu2025towards}                  & \xmark & \cmark & \xmark \\
		SIG~\cite{barni2019new}                     & \cmark & \cmark & \xmark \\
		\midrule
		\rowcolor{blue!10}\textbf{Checkerboard}     & \cmark & \cmark & \cmark \\
		\bottomrule
	\end{tabular}
\end{table}

\paragraph{Learning-Based Poisoning}

Most existing high-performing clean-label attacks fall into this category. These methods rely on surrogate-model training, adversarial perturbation generation, or iterative optimization to strengthen the trigger signal. 

One line of work suppresses benign features to strengthen trigger reliance. LC~\cite{turner2019label} uses adversarial perturbations to construct hard-to-classify label-consistent poisons, while GenBound~\cite{yu2024generalization} formulates poison generation through a clean-label generalization bound. Another line exploits feature or gradient matching. HTBA~\cite{saha2020hidden} aligns poisoned target-class features with trigger-patched source features, whereas SAA~\cite{souri2022sleeper} synthesizes poisons via gradient matching. These methods require auxiliary optimization and are mainly studied in source-specific settings. A third line optimizes triggers directly. Invisible Poison~\cite{ning2021invisible} learns an imperceptible trigger generator, Narcissus~\cite{zeng2023narcissus} optimizes a universal trigger using a surrogate model, and COMBAT~\cite{huynh2024combat} and Hybrid~\cite{yuan2025stealthy} employ joint optimization with generators or auxiliary components. Unlike Checkerboard, their deployed triggers are obtained numerically rather than in closed form. Finally, BAAT~\cite{zhu2025towards} injects semantic attributes using pretrained editors, producing sample-specific label-consistent poisons through semantic rather than norm-bounded perturbations.

\paragraph{Learning-Free Poisoning}

Learning-free attacks apply a fixed trigger pattern without surrogate training or iterative trigger optimization.It is simple and efficient but is harder to make effective under the clean-label constraint because the fixed trigger must compete with benign target-class semantics. SIG~\cite{barni2019new} is a representative static-trigger baseline: it applies a fixed sinusoidal signal to target-class samples without learning. Its trigger is handcrafted rather than derived as the global solution of an explicit analytical trigger-design objective.

Checkerboard targets a largely unexplored region of the clean-label attack design space. Unlike learning-based methods, it requires neither surrogate-model training nor iterative trigger optimization. Unlike existing learning-free approaches, its trigger is derived analytically under standard assumptions on natural-image statistics. The trigger can also be constructed without access to training images and admits a closed-form solution; only target-class samples are required at the poisoning stage. As summarized in Table~\ref{tab:related-properties}, Checkerboard therefore combines learning-free, data-free, and closed-form trigger construction within a single framework.

Beyond these differences in attack construction, checkerboard patterns have previously used as predefined triggers in settings such as dirty-label poisoning, architectural backdoors, and backdoor-defense evaluations~\cite{carlini2022poisoning,xiang2022posttraining,bober2023architectural}. In these works, the pattern is manually specified rather than derived from a trigger-design principle. In contrast, Checkerboard derives the pixel-wise pattern as a closed-form global maximizer of the proposed FDR-based objective under the adopted natural-image prior, providing a principled and analytical basis for its use in low-budget clean-label poisoning.

\section{Method}
\label{sec:method}

\begin{figure}[t]
	\centering
	\includegraphics[width=\columnwidth]{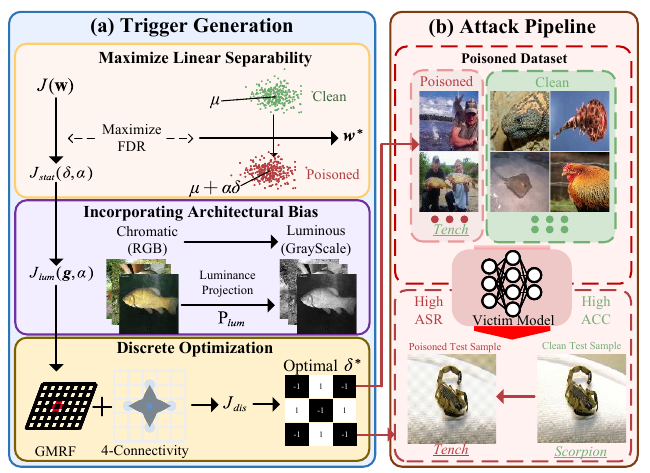}%
	\caption{Overview of Checkerboard.}
	\label{fig:attack-overview}
\end{figure}

We first define the threat model. We then present the design motivation, introduce nature-image prior to derive the closed-form trigger patter solution, and finally describe the full attack pipeline. Figure~\ref{fig:attack-overview} provides an overview of Checkerboard.

\subsection{Threat Model}
\label{sec:threat}

We consider a \textbf{data outsourcing} scenario~\cite{zeng2023narcissus,qiu2024belt,ma2024watch}, where a victim trains a deep model (hereafter referred to as the \emph{victim model}) using a dataset $\mathcal{D}$ collected from multiple external sources. An adversary can manipulate a limited subset of training samples, but cannot modify the victim's training procedure, model architecture, or optimization process.

\subsubsection{Attacker's Goals}

The adversary aims to implant a backdoor into the victim model $f_\theta$ while satisfying three objectives.
\textbf{Backdoor activation:}
Given a trigger direction $\bdelta$ and activation strength $\alpha$, applying the trigger to a high proportion of non-target inputs should cause the victim model to predict an attacker-chosen target class $t$. \textbf{Task preservation:} The victim model should maintain clean classification accuracy close to that obtained through benign training. \textbf{Stealthiness:}
The attack follows the clean-label constraint: poisoned samples preserve their original labels and remain visually compatible with their semantic classes. Moreover, the trigger magnitude is constrained by a small $\ell_\infty$ perturbation budget~\cite{zeng2023narcissus,huynh2024combat}.

\subsubsection{Attacker's Capabilities}

We study a highly restricted attacker model with two components. \textbf{Data-independent trigger design:}
The adversary derives the trigger before accessing any training image. Its construction uses no training samples, no surrogate model, and no numerical optimization. The adversary is only assumed to know that the task is natural-image classification. \textbf{Restricted poison instantiation:}
Let $\mathcal{D}_t:=\{(\bx_i,y_i)\in\mathcal{D}:y_i=t\}$ denote the complete target-class subset. Only after constructing the trigger does the adversary access $\mathcal{D}_t$, but no non-target training samples. The adversary uses $\mathcal{D}_t$ only as a pool of clean images and uniformly selects $m$ distinct samples, where $m\leq|\mathcal{D}_t|$ is the poisoning budget. It applies the already constructed trigger to the selected samples and preserves the original labels. The target-class access is required to create the poisoned examples, but not to derive the trigger. 

These restricted capabilities also make the attack independent of a particular victim architecture or training algorithm during trigger construction. We evaluate this generality across multiple datasets, CNN and Transformer architectures, and both from-scratch training and pretrained-model fine-tuning in Section 4. From a practicality perspective, the attack requires neither auxiliary data nor attack-side model training or iterative optimization, although the required fraction of target-class samples varies across datasets.

\subsection{Design via Distribution Separability}
\label{sec:separability-objective}

\noindent\textbf{Design motivation.}
Clean-label poisoning presents a distinct trigger-design challenge. Because every poisoned image retains its original target label, its natural semantic features already support that label and vary from image to image. The trigger is the attacker-controlled component that can be shared across all poisoned carriers. We therefore seek a trigger pattern that induces a common statistical change that stands out from natural-image variation. Maximizing raw pixel displacement alone is insufficient: under the same norm bound, a shift aligned with a high-variance natural direction can be less distinctive than one aligned with a low-variance direction.

Directly optimizing attack success or shortcut acquisition would depend on the surrogate model and data. Instead, we use an input-space design proxy: the separability between a reference clean-image distribution and the triggered distribution obtained by applying the same fixed pattern to every clean sample. Figure~\ref{fig:separability-intuition} visualizes this design principle: the trigger maps each clean sample to a triggered counterpart, and we seek the bounded direction whose induced distribution shift is largest relative to the natural spread of the clean images. Thus, a separable triggered sample provides a distinctive cue shared across poisoned samples; using this cue only to target-class images makes it statistically correlated with the target label while keeping its construction independent of a particular victim model.

\noindent\textbf{Separability objective.}
Let an RGB image have height $H$, width $W$, $n=HW$ spatial locations, and $d=3n$ scalar coordinates. We denote a clean image as the vector $\bx\in[0,1]^d$ and model it as a random draw $\bx\sim\mathcal{P}_x$ from a reference natural-image distribution with mean $\bmu_x\in\mathbb{R}^d$ and positive-definite covariance $\bSigma_x\in\mathbb{R}^{d\times d}$. This population distribution states the analytical objective, and the attacker neither observes samples from it nor estimates its moments. For a trigger direction $\bdelta\in[-1,1]^d$ and fixed strength $\alpha$, we define the analytically convenient, unclipped triggered sample $\bp\in\mathbb{R}^d$ of $\bx$ as
\begin{equation}
  \bp=\bx+\alpha\bdelta.
  \label{eq:additive-trigger}
\end{equation}
We use $\mathcal{P}_p^{\bdelta,\alpha}$ to denote the distribution of $\bp$ induced by drawing $\bx\sim\mathcal{P}_x$ and applying this same fixed trigger.
Here, $\mathcal{P}_p^{\bdelta,\alpha}$ is a paired design distribution obtained by applying the same candidate trigger to every draw from the reference distribution; it is not intended to reproduce the finite poisoned mixture used for victim training.
We formulate trigger design as
\begin{equation}
  \bdelta^*
  \in\arg\max_{\|\bdelta\|_\infty\le 1}
  \operatorname{Sep}
  (\mathcal{P}_x,\mathcal{P}_p^{\bdelta,\alpha}),
  \label{eq:separability-design}
\end{equation}
where $\operatorname{Sep}$ measures the distribution shift relative to within-distribution variation. The construction therefore selects the direction that maximizes the normalized separation between the clean reference distribution and its triggered counterpart.

We instantiate $\operatorname{Sep}$ using the maximum Fisher discriminant ratio (FDR), which measures the squared projected mean difference normalized by within-distribution variance:
\begin{equation}
  \operatorname{Sep}_{\mathrm{FDR}}
  (\mathcal{P}_x,\mathcal{P}_p^{\bdelta,\alpha})
  =\max_{\bm{w}\ne 0}
  \frac{\bigl(\mathbb{E}[\bm{w}^{\top}\bp]
        -\mathbb{E}[\bm{w}^{\top}\bx]\bigr)^2}
       {\operatorname{Var}(\bm{w}^{\top}\bp)
        +\operatorname{Var}(\bm{w}^{\top}\bx)}.
  \label{eq:fdr-definition}
\end{equation}
The projection $\bm{w}$ measures the most discriminative direction, and maximizing over $\bm{w}$ yields a scalar criterion that favors trigger shifts along low-variance natural-image directions. Under Equation~\eqref{eq:additive-trigger}, the triggered distribution has mean $\bmu_p=\bmu_x+\alpha\bdelta$ and covariance $\bSigma_p=\bSigma_x$, which directly matches the design motivation. Substituting these into Equation~\eqref{eq:fdr-definition} and maximizing the resulting generalized Rayleigh quotient gives
\begin{equation}
  \operatorname{Sep}_{\mathrm{FDR}}
  (\mathcal{P}_x,\mathcal{P}_p^{\bdelta,\alpha})
  =\frac{\alpha^2}{2}
   \bdelta^{\top}\bSigma_x^{-1}\bdelta,
  \qquad
  \bm{w}^{*}\propto\bSigma_x^{-1}\bdelta
  \quad(\bdelta\ne\bm{0}).
  \label{eq:fdr-unclipped}
\end{equation}
Thus, for fixed $\alpha$, maximizing the clean-to-triggered FDR is equivalent to maximizing the Mahalanobis magnitude of the trigger relative to natural-image variation:
\begin{equation}
  \max_{\|\bdelta\|_\infty\le 1}
  \bdelta^\top\bSigma_x^{-1}\bdelta.
  \label{eq:ideal-mahalanobis}
\end{equation}
Appendix~\ref{app:theory} provides the full derivation and separates two questions. The \emph{construction} problem identifies the bounded fixed pattern that maximizes the normalized clean-to-triggered distribution shift under a model-agnostic natural-image prior, as characterized by Equations~\eqref{eq:fdr-unclipped}--\eqref{eq:ideal-mahalanobis}. The \emph{acquisition} problem asks whether victim training learns to associate this shared pattern with the target label. This depends on the poison budget and training dynamics and is evaluated empirically in Section~\ref{sec:evaluation} using matched trigger-family comparisons.

\subsection{A Data-Independent Natural-Image Prior}
\label{sec:graph-prior}

\noindent\textbf{Luminance restriction.}
Equation~\eqref{eq:ideal-mahalanobis} is defined directly in RGB space and may favor low-variance chromatic directions. Although statistically distinctive, such directions need not provide a stable spatial cue across vision architectures. Natural images are dominated by low-frequency luminance structure~\cite{Field:87,ruderman1998statistics}, making channel-shared luminance contrast a broadly applicable image-space cue. We therefore restrict the trigger to a scalar, channel-shared luminance pattern $\bg\in[-1,1]^n$. Specifically, the fixed replication operator is defined by $R\bg=(\bg^\top,\bg^\top,\bg^\top)^\top$, so $\bdelta=R\bg$. The induced luminance-space problem is
\begin{equation}
  \max_{\bg\in[-1,1]^n}
  \bg^\top\bQ_{\mathrm{lum}}\bg,
  \qquad
  \bQ_{\mathrm{lum}}:=R^\top\bSigma_x^{-1}R.
  \label{eq:luminance-objective}
\end{equation}
The problem therefore reduces to finding the channel-shared luminance pattern that maximizes this induced precision-weighted energy. Directly solving Equation~\eqref{eq:luminance-objective}, however, requires estimating $\bQ_{\mathrm{lum}}$ and clean-image precision $\bSigma_x^{-1}$, which is unavailable to the attacker under our data-independent construction. Such estimation can also be unstable in high-dimensional image space: a sample covariance may be poorly conditioned or singular, and inversion magnifies errors associated with small eigenvalues~\cite{qu2003building,ledoit2004well}. We therefore replace empirical precision estimation with two explicit structural assumptions about natural images.

\noindent\textbf{A1. Local-smoothness precision model.}
Natural images exhibit strong local correlations and predominantly low-frequency structure~\cite{hyvarinen2009natural,Field:87,ruderman1994statistics}. Following a local Gaussian Markov random-field model~\cite{rue2005gaussian}, we approximate the induced luminance precision by a positively weighted graph Laplacian with a ridge term:
\begin{equation*}
  \bQ_{\mathrm{lum}}
  \approx \lambda L_w+\beta I,
  \qquad \lambda>0,\;\beta>0.
\end{equation*}
The Laplacian component represents first-order local smoothness, and the ridge is introduced only to make the surrogate positive definite. This is a generic reference prior, not an estimate of the attacker's particular target class.

\noindent\textbf{A2. Four-neighbor image lattice.}
We instantiate $L_w$ on the ordinary, non-wrapping, connected four-neighbor image grid $G=(V,E)$, where each pixel is connected to its horizontal and vertical neighbors and every undirected edge is counted once. The deployed construction uses unit weights, $w_{ij}=1$. We retain general fixed weights $w_{ij}>0$ below because Appendix~\ref{app:graph-sensitivity} shows that any positive reweighting preserves the derived pattern. The resulting trigger-design problem is
\begin{align}
  \max_{\bg\in[-1,1]^n}\quad
  \widetilde{a}(\bg)
  &=\bg^\top(\lambda L_w+\beta I)\bg \nonumber\\
  &=\lambda\sum_{(i,j)\in E}w_{ij}(g_i-g_j)^2
    +\beta\sum_{i\in V}g_i^2.
  \label{eq:graph-objective}
\end{align}

Under this precision surrogate, locally smooth directions have low graph energy, whereas patterns with strong adjacent-pixel contrast have high graph energy. Maximizing the FDR proxy therefore favors spatial alternation. For every $\lambda>0$, $\beta>0$, and set of positive four-neighbor edge weights, Equation~\eqref{eq:graph-objective} has exactly two global maximizers:
\begin{equation}
  g^*_{u,v}=(-1)^{u+v}
  \quad\text{and}\quad -g^*_{u,v}.
  \label{eq:checkerboard}
\end{equation}
Of the sign-equivalent phases, we use $g^*_{u,v}=(-1)^{u+v}$ for zero-based $u,v$, replicated across RGB for poisoning and activation. We adopt the four-neighbor grid because it is the minimal undirected first-order adjacency that treats horizontal and vertical neighbors symmetrically, and its maximizer requires no attack-side model, image data, or tuned relative-neighborhood parameters. Appendix~\ref{app:graph-sensitivity} proves the solution, shows its invariance to positive $\lambda$, $\beta$, and edge weights, and analyzes the patterns induced by alternative priors. The four-neighbor structure is therefore a data-independent design prior compatible with our attacker constraints.

\subsection{Clipping and Attack Pipeline}
\label{sec:attack-construction}

The derivation above uses the unclipped additive model for analytical convenience. Let $\bdelta^*=R\bg^*$ denote the unit-amplitude checkerboard direction. For an image $\bx\in[0,1]^d$ and a trigger strength $\alpha\geq0$, we define the clipped trigger operator
\begin{equation}
  T_{\alpha}(\bx)
  =\operatorname{clip}(\bx+\alpha\bdelta^*,0,1),
  \label{eq:clipped-poison}
\end{equation}
which guarantees $\|T_\rho(\bx)-\bx\|_\infty\leq\alpha$. Clipping makes the exact mean displacement and pooled covariance depend on both the clean-image distribution and candidate pattern, so the exact clipped optimizer is distribution-dependent in general. Appendix~\ref{app:clipping} gives this exact formulation and then analyzes a clipping-aware counterpart of the adopted prior. Under a common coordinatewise monotone clipping response and a fixed positively weighted four-neighbor model, the two checkerboard phases maximize the resulting clipping-aware contrast objective. Thus, clipping changes the attainable contrast magnitude but not the pattern selected by this conditional model.

The complete attack pipeline consists of three steps:
\begin{enumerate}
  \item \textbf{Construct the trigger.} Set $g^*_{u,v}=(-1)^{u+v}$ and replicate it across channels once to obtain the direction $\bdelta^*=R\bg^*$.
  \item \textbf{Replace target-class samples.} Uniformly sample a training set $\mathcal{S}\subseteq\mathcal{D}_t$ with $|\mathcal{S}|=m$. For every $(\bx,y_t)\in\mathcal{S}$, replace $(\bx,y_t)$ with $(T_\alpha(\bx),y_t)$, thereby preserving its original target-class label.
  \item \textbf{Activate the backdoor.} At inference, the attacker applies $T_{\gamma \alpha}$ to a non-target input under its control. The native setting uses the native strength $\gamma=1$; larger $\gamma$ values are explicitly labeled as stronger activation.  During inference, the adversary activates the backdoor by applying the trigger to a clean test input $\bx_{test}$. Following established practices in clean-label attacks~\cite{zeng2023narcissus}, we use test-time trigger amplification. The perturbed test sample is formulated as: $\bp_{test} = T_{\gamma\alpha}(\bx_{test})=\text{clip}(\bx_{test} + \gamma \alpha \bdelta^*,0,1)$
  where $\gamma \ge 1$ is an amplification factor. Consistent with previous studied~\cite{zeng2023narcissus}, we find that it can effectively boost attack performance when amplification is used.
\end{enumerate}

The proposed Checkerboard attack has a closed-form and analytical trigger and does not require surrogate-model training or numerical optimization. Constructing $\bdelta^*$ costs $O(d)$ once, generating $m$ poisoned samples costs $O(md)$, and applying the trigger to one test input costs $O(d)$, with constant additional memory per image. Therefore, our attack is simple, scalable, and computationally efficient.

\section{Evaluation}
\label{sec:evaluation}

We evaluate Checkerboard from two main perspectives:
\begin{itemize}
	\item \textbf{Attack performance and efficiency.} We compare Checkerboard with representative clean-label backdoor attacks in terms of attack success, clean accuracy, and poison-generation cost across four benchmark datasets.
	\item \textbf{Coverage and sensitivity.} We evaluate Checkerboard across diverse model architectures under both from-scratch training and fine-tuning, and further examine its sensitivity to poison count, trigger strength, target class, and trigger-pattern construction.

\end{itemize}

\subsection{Experimental Setup}
\label{sec:setup}

\paragraph{Datasets and budgets.}
We use CIFAR-10 and CIFAR-100~\cite{krizhevsky2009learning}, CelebA~\cite{liu2015deep}, and IN-100. Following prior work~\cite{nguyen2021wanet,huynh2024combat}, we form CelebA-8 from the binary attributes \emph{Heavy Makeup}, \emph{Mouth Slightly Open}, and \emph{Smiling}. Let $N$ be the training-set size, $N_t$ the number of target-class training samples, and $m$ the number of poisoned samples. IN-100 is constructed from the first 100 classes sorted alphabetically of ImageNet-1k~\cite{deng2009imagenet}. Table~\ref{tab:dataset-budget} provides a summary of the primary native-activation budgets and reports both global ($m/N$) and target-class ($m/N_t$) poison rates.

\begin{table}[t]
	\centering
	\caption{Dataset statistics summary. $N$ and $N_t$ denote the number of samples in the dataset and the number of samples in the target class, respectively.}
	\label{tab:dataset-budget}
	\small
	\setlength{\tabcolsep}{4pt}
	\begin{tabular}{lcccc}
		\toprule
		& CIFAR-10 & CIFAR-100 & CelebA-8 & IN-100 \\
		\midrule
		Classes
		& 10 & 100 & 8 & 100 \\
		
		Resolution
		& $32\!\times\!32$
		& $32\!\times\!32$
		& $64\!\times\!64$
		& $224\!\times\!224$ \\
		
		$N$
		& 50,000
		& 50,000
		& 162,770
		& 129,395 \\
		
		$N_t$
		& 5,000
		& 500
		& 46,088
		& 1,300 \\
		
		$m$
		& 20
		& 50
		& 20
		& 520 \\
		
		$m/N$
		& 0.04\%
		& 0.1\%
		& 0.0123\%
		& 0.402\% \\
		
		$m/N_t$
		& 0.4\%
		& 10\%
		& 0.0434\%
		& 40\% \\
		
		$\alpha$
		& $10/255$
		& $10/255$
		& $10/255$
		& $16/255$ \\
		\bottomrule
	\end{tabular}
\end{table}

\paragraph{Victim training.}
Unless otherwise specified, the victim is ResNet-18~\cite{he2016deep}, trained with SGD, momentum 0.9, batch size 256, cosine learning-rate decay from 0.1, and the dataset-specific epoch count, weight decay, augmentation, and normalization recorded in Appendix~\ref{app:experimental-details}.

\paragraph{Targets, repetitions, and metrics.}
Unless otherwise noted, we fix the target class to 0. We report two standard metrics: \textbf{clean accuracy (ACC)}, i.e., the classification accuracy on clean test inputs, and \textbf{attack success rate (ASR)}, i.e., the probability that triggered test samples are classified as the target class. All reported results are averaged over five runs.

\begin{table*}[t]
	\centering
	\caption{Performance comparison of clean-label backdoor attacks. Bold marks the highest completed ASR among the directly comparable norm-bounded attacks in each column. }
	\label{tab:main-attack-results}
	\scriptsize
	\setlength{\tabcolsep}{2.8pt}
	\resizebox{\textwidth}{!}{%
		\begin{tabular}{lccc ccc ccc ccc}
			\toprule
			& \multicolumn{3}{c}{CIFAR-10} & \multicolumn{3}{c}{CIFAR-100} & \multicolumn{3}{c}{CelebA-8} & \multicolumn{3}{c}{IN-100} \\
			\cmidrule(lr){2-4}\cmidrule(lr){5-7}\cmidrule(lr){8-10}\cmidrule(lr){11-13}
			Attack
			& ACC & \makecell{ASR\\$\gamma=1$} & \makecell{ASR\\$\gamma=1.5$}
			& ACC & \makecell{ASR\\$\gamma=1$} & \makecell{ASR\\$\gamma=1.5$}
			& ACC & \makecell{ASR\\$\gamma=1$} & \makecell{ASR\\$\gamma=1.5$}
			& ACC & \makecell{ASR\\$\gamma=1$} & \makecell{ASR\\$\gamma=1.5$} \\
			\midrule
			Clean (unpoisoned)
			& 94.14 & 5.20 & 5.18
			& 76.78 & 0.07 & 0.08
			& 79.49 & 0.01 & 0.01
			& 70.34 & 0.01 & 0.01 \\
			\midrule
			\multicolumn{13}{l}{\emph{Norm-bounded attacks}}\\
			LC
			& 94.32 & 0.57 & 0.52
			& 76.93 & 0.15 & 0.15
			& 79.17 & 6.11 & 6.65
			& 70.35 & 0.13 & 0.15 \\
			SIG
			& 94.72 & 0.36 & 0.37
			& 76.56 & 0.35 & 0.36
			& 79.06 & 7.17 & 8.84
			& 69.93 & 0.11 & 0.11 \\
			Invisible Poison
			& 94.60 & 0.60 & 0.63
			& 74.80 & 9.03 & 25.40
			& 80.40 & 29.06 & 36.45
			& 70.17 & 0.12 & 0.21 \\
			Narcissus
			& 95.25 & 0.53 & 0.97
			& 76.75 & 11.42 & 25.48
			& 79.87 & 8.29 & 9.04
			& 70.79 & 10.25 & 19.23 \\
			GenBound
			& 94.60 & 10.18 & 11.67
			& 76.92 & 8.04 & 10.08
			& 80.27 & 47.82 & 47.69
			& 70.22 & 1.37 & 2.98 \\
			Hybrid
			& 95.20 & 7.04 & 8.15
			& 76.68 & 0.20 & 0.35
			& 79.91 & 42.09 & 66.80
			& 70.70 & 0.15 & 0.21 \\
			COMBAT
			& 94.98 & 6.08 & 23.17
			& 74.63 & 4.77 & 13.65
			& 78.40 & 36.45 & 53.35
			& 68.91 & 57.68 & 85.43 \\
			\addlinespace[1pt]
			\textbf{Checkerboard}
			& 94.55 & \textbf{95.72} & \textbf{99.76}
			& 76.66 & \textbf{96.39} & \textbf{99.48}
			& 79.24 & \textbf{100.00} & \textbf{100.00}
			& 69.94 & \textbf{83.76} & \textbf{98.14} \\
			\midrule
			\multicolumn{13}{l}{\emph{Different attack scope or perturbation model}}\\
			SAA (source-specific)
			& 94.65 & 0.70 & 0.73
			& 75.82 & 1.00 & 1.03
			& 78.24 & 36.10 & 40.32
			& 68.80 & 0.00 & 0.00 \\
			BAAT (semantic edit)
			& 94.15 & 34.99 & N/A
			& 76.95 & 83.30 & N/A
			& 79.60 & 37.01 & N/A
			& 69.02 & 93.40 & N/A \\
			\bottomrule
		\end{tabular}%
	}
\end{table*}

\subsection{Attack Performance}
\label{sec:attack-performance}

We compare Checkerboard against nine representative clean-label backdoor attacks: LC~\cite{turner2019label}, SIG~\cite{barni2019new}, SAA~\cite{souri2022sleeper}, Narcissus~\cite{zeng2023narcissus}, Invisible Poison~\cite{ning2021invisible}, GenBound~\cite{yu2024generalization}, Hybrid~\cite{yuan2025stealthy}, COMBAT~\cite{huynh2024combat}, and BAAT~\cite{zhu2025towards}. All baselines are implemented using their official codebases or published specifications. To ensure a fair comparison, all attacks except BAAT are evaluated under the same training-time \(\ell_\infty\) perturbation budget whenever applicable. BAAT instead uses a generative model to produce poisoned samples by modifying semantic attributes of clean inputs, and its perturbation is therefore not naturally characterized by an \(\ell_\infty\) bound.

For baselines that require a surrogate for poison generation, we use a ResNet-18 surrogate matching the victim architecture, which provides these methods with a favorable white-box setting. SAA is a one-to-one attack; following its original threat model, we set class 1 as the source class and compute ASR only on triggered source-class test samples. All other methods, including Checkerboard, are evaluated in the all-to-one setting, with target class set to 0 and ASR computed over all non-target test samples.

Table~\ref{tab:main-attack-results} summarizes the overall comparison. The \(\gamma=1\) and \(1.5\) columns report ASR when the inference-time trigger perturbation matches the training-time perturbation and is amplified by \(1.5\times\), respectively. Notably, the clean (unpoisoned) models exhibit at most \(5.20\%\) ASR when evaluated with the same Checkerboard trigger, indicating that the high ASR of Checkerboard is not a pre-existing model response but is learned through poisoned training. Checkerboard achieves substantially higher ASR than the norm-bounded clean-label baselines while maintaining clean accuracy close to that of the clean model. In contrast, under the same training-time perturbation bound and dataset-specific poisoning budgets in Table~\ref{tab:dataset-budget}, many existing baselines remain near zero ASR, especially on CIFAR-10, CIFAR-100, and IN-100.


On CIFAR-10, the limitations of existing clean-label attacks become apparent under these constrained settings. Although BAAT achieves the strongest baseline ASR at 34.99\%, most other methods fail to exceed 1\%. In comparison, Checkerboard ($\gamma =  1$) already reaches 95.72\% ASR and further improves to a near-perfect 99.76\% under a moderate $\times 1.5$ test-time amplification.

CIFAR-100 is a difficult dataset because its 100 classes contain only 500 training samples each, which limits the number of poisoned samples allowed under a fixed budget. Among the norm-bounded baselines, Narcissus only reaches 25.48\% ASR even under a $\times 1.5$ trigger amplification. BAAT also performs strongly, achieving 83.3\% ASR, although it is not directly comparable because it does not operate under the same training-time \(\ell_\infty\) perturbation constraint. In contrast, Checkerboard achieves 96.39\% ASR natively at $\gamma= 1$ and 99.48\% at $\gamma= 1.5$, demonstrating strong effectiveness and sample efficiency under a strict bounded-perturbation setting.

Checkerboard exhibits its strongest performance on CelebA, achieving 100\% ASR at \(\gamma = 1\). Compared with the other datasets, CelebA consists of aligned human faces that typically contain smoother and lower-frequency background. We hypothesize that this increases the contrast between the Checkerboard trigger and the background, making the trigger easier for the model to learn.

\begin{figure}[t]
	\centering
	\captionsetup[subfigure]{font=scriptsize}
	\begin{subfigure}[t]{0.235\columnwidth}
		\centering
		\includegraphics[width=\linewidth]{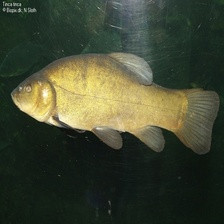}
		\caption{Clean}
	\end{subfigure}\hfill
	\begin{subfigure}[t]{0.235\columnwidth}
		\centering
		\includegraphics[width=\linewidth]{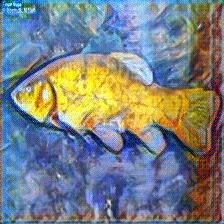}
		\caption{BAAT}
	\end{subfigure}\hfill
	\begin{subfigure}[t]{0.235\columnwidth}
		\centering
		\includegraphics[width=\linewidth]{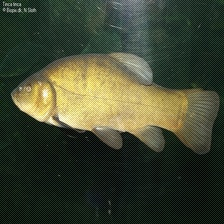}%
		\caption{CB $16/255$}
	\end{subfigure}\hfill
	\begin{subfigure}[t]{0.235\columnwidth}
		\centering
		\includegraphics[width=\linewidth]{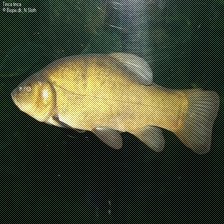}%
		\caption{CB $24/255$}
	\end{subfigure}
	\caption{Samples of clean and poisoned samples generated by BAAT and Checkerboard on IN-100. (c) and (d) shows the Checkerboard poisoned samples when trigger perturbations are $16/255 (\gamma=1)$ and $24/255 (\gamma=1.5)$, respectively.}
	\label{fig:in100-baat-comparison}
\end{figure}

IN-100 is the most challenging dataset because of its larger input resolution, and richer texture variations. Nevertheless, Checkerboard ($\gamma= 1$) achieves an ASR of 83.76\%, substantially outperforming the strongest norm-bounded baseline under the same \(16/255\) perturbation budget. With $\times 1.5$ trigger amplification, the ASR further increases to 98.14\%, showing that the attack remains highly effective even in texture-rich, high-resolution settings. BAAT also achieves strong ASR on IN-100, but its poisoned samples deviate much more from their clean counterparts than ours, as shown in Figure~\ref{fig:in100-baat-comparison}.

For completeness, Appendix~\ref{app:cifar10-fixed500} reports a separate CIFAR-10 comparison using the published settings of each baseline.

\subsection{Poison-Generation Efficiency}
\label{sec:efficiency}

\begin{figure}[t]
	\centering
	\includegraphics[width=\columnwidth]{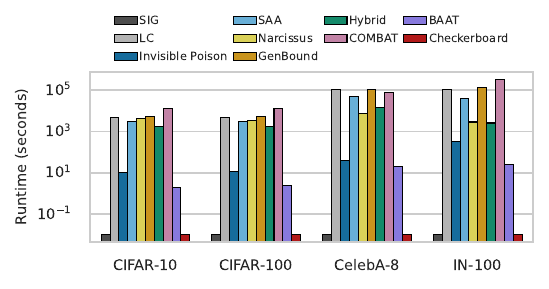}%
	\caption{Poison-generation time (logarithmic scale).}
	\label{fig:runtime}
\end{figure}

Figure~\ref{fig:runtime} compares poison-generation cost. For BAAT, the displayed value excludes generative-model pretraining and therefore understates its end-to-end setup cost, making it not directly comparable to methods whose full attack-specific computation is included.

The surrogate- or optimization-intensive methods LC, SAA, Narcissus, GenBound, Hybrid, and COMBAT generally require $10^3$--$10^5$ seconds across the evaluated datasets. On IN-100, LC, GenBound, and COMBAT reach or exceed $10^5$ seconds. Checkerboard instead constructs its trigger analytically and directly adds it to uniformly sampled target-class images. Its measured poison-generation time remains below $0.1$ seconds on all datasets, placing it in the same sub-$0.1$-second regime as SIG and orders of magnitude below the optimization-based baselines. These results attribute Checkerboard's low generation cost to closed-form trigger construction and direct poison injection rather than learned or iterative trigger pattern synthesis.

\subsection{Victim and Training-Regime Coverage}
\label{sec:coverage}

\subsubsection{From-Scratch Architectures}
\label{sec:architecture-coverage}

To evaluate the robustness of Checkerboard across model families, we train ResNet-18, VGG16, MobileNetV3, DenseNet-121, and 2 Transformers CCT and ViT-Small from scratch on CIFAR-10 under the same protocol summarized in Table~\ref{tab:model-comparison}.

The results show that Checkerboard remains effective across diverse victim architectures, consistent with its separability-based design. Under the \(m=20\) setting, models with stronger clean-task performance also tend to achieve higher native ASR. For example, DenseNet-121 achieves both the highest clean ACC (\(95.53\%\)) and the highest native ASR (\(98.83\%\)). By contrast, CCT and ViT-Small achieve lower native ASRs of \(50.23\%\) and \(35.77\%\), respectively. Increasing the poison count to \(m=100\) raises their native ASRs to \(76.06\%\) and \(53.58\%\); with 2 times test-time amplification, they further reach \(99.89\%\) and \(94.58\%\). Overall, these results demonstrate Checkerboard’s effectiveness across both CNN and Transformer architectures, while the evaluated Transformer models require more poisoned samples or stronger test-time activation to attain ASRs comparable to those of the CNNs. One plausible explanation is that Transformers are generally more data-demanding when trained from scratch, whereas CIFAR-10 provides only 50,000 low-resolution training images. This may limit both their clean-task learning and their acquisition of the trigger–target association. Consistent with this explanation, ACC and native ASR follow a similar trend across the evaluated architectures: models that learn the clean task more effectively also tend to acquire the backdoor more strongly.

\begin{table}[t]
	\centering
	\caption{From-scratch victim-architecture results on CIFAR-10, target 0, and $\alpha=10/255$.}
	\label{tab:model-comparison}
	\resizebox{\columnwidth}{!}{%
		\begin{tabular}{cccccc}
			\toprule
			\multirow[c]{2}{*}[-0.8ex]{Model}
			& \multirow[c]{2}{*}[-0.8ex]{$m$}
			& \multirow[c]{2}{*}[-0.8ex]{ACC}
			& \multicolumn{1}{c}{Native ASR}
			& \multicolumn{2}{c}{Amplified ASR} \\
			\cmidrule(lr){4-4}\cmidrule(lr){5-6}
			& & &
			\makecell{$10/255$ \\ ($\gamma=1$)}
			& \makecell{$15/255$ \\ ($\gamma=1.5$)}
			& \makecell{$20/255$ \\ ($\gamma=2$)} \\
			\midrule
			ResNet-18~\cite{he2016deep}    & 20  & 94.55 & 95.72 & 99.76 & 99.99 \\
			VGG16~\cite{simonyan2014very}        & 20  & 93.06 & 92.13 & 99.40 & 99.94 \\
			MobileNetV3~\cite{howard2019searching}  & 20  & 92.67 & 89.09 & 97.64 & 98.76 \\
			DenseNet-121~\cite{huang2017densely} & 20  & 95.53 & 98.83 & 99.90 & 99.99 \\
			\cmidrule(lr){2-6}
			\multirow[c]{2}{*}{CCT~\cite{hassani2021escaping}}
			             & 20  & 88.35 & 50.23 & 75.23 & 89.26 \\
			
			             & 100 & 88.38 & 76.06 & 98.37 & 99.89 \\
			\addlinespace[1pt]
			\cmidrule(lr){2-6}
			\multirow[c]{2}{*}{ViT-Small~\cite{dosovitskiy2020image}}
			             & 20  & 83.81 & 35.77 & 58.73 & 79.63 \\
			
			             & 100 & 83.72 & 53.58 & 84.18 & 94.58 \\
			\bottomrule
		\end{tabular}%
	}
\end{table}

\begin{table}[t]
	\centering
	\caption{IN-100 full-model fine-tuning using ImageNet-1k-pretrained checkpoints, target 0, $m=520$, and $\alpha=16/255$.}
	\label{tab:finetune-activation}
	\resizebox{\columnwidth}{!}{%
		\begin{tabular}{ccccc}
			\toprule
			\multirow[c]{2}{*}[-0.8ex]{Model}
			& \multirow[c]{2}{*}[-0.8ex]{ACC}
			& \multicolumn{1}{c}{Native ASR}
			& \multicolumn{2}{c}{Amplified ASR} \\
			\cmidrule(lr){3-3}\cmidrule(lr){4-5}
			&
			& \makecell{$16/255$ \\ ($\gamma=1$)}
			& \makecell{$20/255$ \\ ($\gamma=1.25$)}
			& \makecell{$24/255$ \\ ($\gamma=1.5$)} \\
			\midrule
			ResNet-50  & 80.85 & 82.13 & 92.47 & 96.93 \\
			ConvNeXt-S & 83.19 & 84.51 & 93.29 & 97.68 \\
			Swin-T     & 80.98 & 79.21 & 86.76 & 91.56 \\
			\bottomrule
		\end{tabular}%
	}
\end{table}

\subsubsection{Pretrained-Model Fine-Tuning}
\label{sec:finetuning}

The preceding architecture study isolates breadth under from-scratch CIFAR-10 training. We next evaluate a practically distinct high-resolution regime by loading ImageNet-1k-pretrained ResNet-50, ConvNeXt-S, and Swin-T checkpoints and fine-tuning all model parameters on IN-100. We use the same poisoning configurations as in Section~\ref{sec:attack-performance}. The ACC and ASR columns are obtained from the corresponding poisoned-model fine-tuning runs. 

The experimental results are reported in Table~\ref{tab:finetune-activation}. Under native strength ($\gamma=1$), ASR is $82.13\%$ for ResNet-50, $84.51\%$ for ConvNeXt-S, and $79.21\%$ for Swin-T, while their ACC ranges from $80.85\%$ to $83.19\%$. Thus, Checkerboard is acquired by all backbones after full-model fine-tuning. Increasing only the test-time amplification raises ASR for every backbone: at $\gamma=1.25$, ASR ranges from $86.76\%$ to $93.29\%$, and at $\gamma=1.5$, it ranges from $91.56\%$ to $97.68\%$.  Overall, the results suggest that Checkerboard is also effective in fine-tuning attacks on modern vision classification architectures.

\subsection{Attack Sensitivity}
\label{sec:attack-sensitivity}

\subsubsection{Poison-Count and Trigger-Strength Sensitivity}
\label{sec:budget-sensitivity}

\begin{figure}[t]
  \centering
  \begin{subfigure}[t]{0.48\columnwidth}
    \centering
    \includegraphics[width=\linewidth]{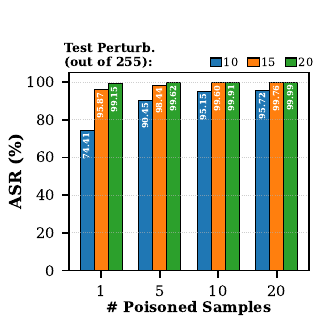}
    \caption{CIFAR-10: $m$ sweep.}
    \label{fig:budget-cifar-count}
  \end{subfigure}\hfill
  \begin{subfigure}[t]{0.48\columnwidth}
    \centering
    \includegraphics[width=\linewidth]{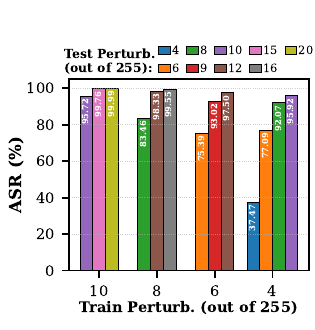}
    \caption{CIFAR-10: $\alpha$ sweep.}
    \label{fig:budget-cifar-strength}
  \end{subfigure}

  \begin{subfigure}[t]{0.48\columnwidth}
    \centering
     \includegraphics[width=\linewidth]{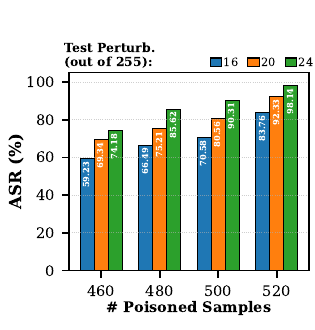}%
    \caption{IN-100: $m$ sweep.}
    \label{fig:budget-in100-count}
  \end{subfigure}\hfill
  \begin{subfigure}[t]{0.48\columnwidth}
    \centering
     \includegraphics[width=\linewidth]{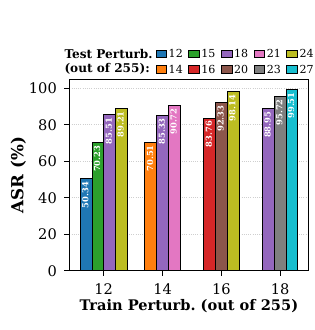}%
    \caption{IN-100: $\alpha$ sweep.}
    \label{fig:budget-in100-strength}
  \end{subfigure}
  \caption{Sensitivity to poison count and trigger strength. Panels (a) and (c) vary $m$ at fixed $\alpha=10/255$ and $16/255$; panels (b) and (d) vary $\alpha$ at fixed $m=20$ and $520$.}
  \label{fig:budget-sensitivity}
\end{figure}

We investigate how the number of poisoned samples ($m$), training-time trigger perturbation ($\alpha$), and test-time trigger amplification ($\gamma$) affect Checkerboard on CIFAR-10 and IN-100, with results shown in Figure~\ref{fig:budget-sensitivity}. Overall, the attack becomes stronger as the number of poisoned samples or the trigger magnitude increases. Nevertheless, Checkerboard can remain effective under constrained training settings when a stronger trigger is applied at test time.

On CIFAR-10, Figure~\ref{fig:budget-sensitivity}(a) shows that Checkerboard remains effective even when very few poisoned samples are available. With $\alpha=10/255$, poisoning only one training sample achieves $74.41\%$ ASR under native activation ($\gamma=1$); increasing the test-time strength to $15/255$ raises ASR to $95.87\%$. Figure~\ref{fig:budget-sensitivity}(b) further shows the effect of reducing the training-time strength. With $m=20$ and $\alpha=4/255$, native ASR is $37.47\%$, whereas increasing the test-time perturbation to $10/255$ raises ASR to $95.92\%$.

The IN-100 results exhibit a similar trend. As shown in Figure~\ref{fig:budget-sensitivity}(c), with $\alpha=16/255$, increasing $m$ from $460$ to $520$ raises native ASR from $59.23\%$ to $83.76\%$; under the stronger test-time strength $24/255$, ASR increases from $74.18\%$ to $98.14\%$. Figure~\ref{fig:budget-sensitivity}(d) shows that, with $m=520$, reducing $\alpha$ from $18/255$ to $12/255$ lowers native ASR from $88.95\%$ to $50.34\%$. However, for $\alpha=12/255$, increasing the test-time strength to $24/255$ raises ASR to $89.21\%$. These results demonstrate that both the poison count and training-time trigger strength influence attack effectiveness, while stronger test-time activation can substantially increase ASR under otherwise constrained training configurations.

\subsubsection{Target-Class Sensitivity}
\label{sec:target-sensitivity}

To measure target dependence, we evaluate Checkerboard on all ten CIFAR-10 targets, all eight CelebA-8 targets, and targets 0--9 for both CIFAR-100 and IN-100. Table~\ref{tab:multi-target} reports the mean across the evaluated targets for each dataset and the corresponding per-target results are in Appendix~\ref{app:multi-target}.

\begin{table}[t]
  \centering
  \caption{Checkerboard target-class sensitivity across evaluated targets under the dataset-specific settings in Table~\ref{tab:dataset-budget}.}
  \label{tab:multi-target}
  \small
  \begin{tabular}{lrrcc}
    \toprule
    Dataset & No. targets & $m$ & ACC & Native ASR \\
    \midrule
    CIFAR-10  & 10 & 20  & 94.43 & 93.04 \\
    CIFAR-100 & 10 & 50  & 76.60 & 97.60 \\
    CelebA-8  & 8  & 20  & 79.27 & 100.00 \\
    IN-100    & 10 & 520 & 69.85 & 82.92 \\
    \bottomrule
  \end{tabular}
\end{table}

Across the complete target sets of CIFAR-10 and CelebA-8, Checkerboard achieves mean native ASRs of $93.04\%$ and $100.00\%$, respectively. For targets 0--9 on CIFAR-100 and IN-100, the corresponding means are $97.60\%$ and $82.92\%$. The mean ACC values are $94.43\%$, $76.60\%$, $79.27\%$, and $69.85\%$ on CIFAR-10, CIFAR-100, CelebA-8, and IN-100, respectively. Thus, the target-0 results are not isolated outcomes within the evaluated target sets.

\subsection{Matched Trigger-Construction Study}
\label{sec:pattern-study}

The analysis in Section~\ref{sec:graph-prior} derives the pixel checkerboard under the four-neighbor prior, but it does not establish that this pattern attains the highest ASR after victim training. We therefore use a matched study to isolate the effects of empirical versus structural precision, unrestricted versus channel-shared perturbations, neighborhood topology, and spatial scale on trigger acquisition by the victim. We perform this study on CIFAR-10 because its RGB dimension permits computationally tractable and direct empirical precision estimation.

\begin{figure}[t]
	\centering
	
	\begin{subfigure}[t]{0.24\columnwidth}
		\centering
		\includegraphics[width=1\columnwidth]{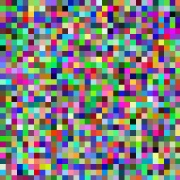}
		\caption{Random}
		\label{fig:pattern-random}
	\end{subfigure}
	\hfill
	\begin{subfigure}[t]{0.24\columnwidth}
		\centering
		\includegraphics[width=1\columnwidth]{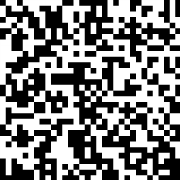}
		\caption{S\&P}
		\label{fig:pattern-sp}
	\end{subfigure}
	\hfill
	\begin{subfigure}[t]{0.24\columnwidth}
		\centering
		\includegraphics[width=1\columnwidth]{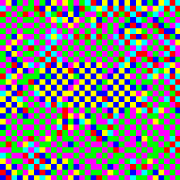}
		\caption{Eq.~\eqref{eq:ideal-mahalanobis}}
		\label{fig:pattern-mahalanobis}
	\end{subfigure}
	\hfill
	\begin{subfigure}[t]{0.24\columnwidth}
		\centering
		\includegraphics[width=1\columnwidth]{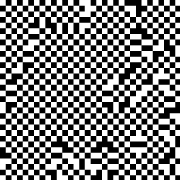}
		\caption{Eq.~\eqref{eq:luminance-objective}}
		\label{fig:pattern-luminance}
	\end{subfigure}
	
	\vspace{1mm}
	
	\begin{subfigure}[t]{0.24\columnwidth}
		\centering
		\includegraphics[width=1\columnwidth]{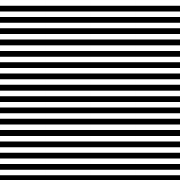}
		\caption{H-stripes}
		\label{fig:pattern-hstripes}
	\end{subfigure}
	\hfill
	\begin{subfigure}[t]{0.24\columnwidth}
		\centering
		\includegraphics[width=1\columnwidth]{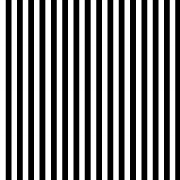}
		\caption{V-stripes}
		\label{fig:pattern-vstripes}
	\end{subfigure}
	\hfill
	\begin{subfigure}[t]{0.24\columnwidth}
		\centering
		\includegraphics[width=1\columnwidth]{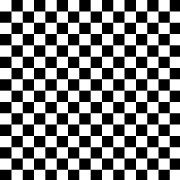}
		\caption{$2\!\times\!2$ block}
		\label{fig:pattern-block}
	\end{subfigure}
	\hfill
	\begin{subfigure}[t]{0.24\columnwidth}
		\centering
		\includegraphics[width=1\columnwidth]{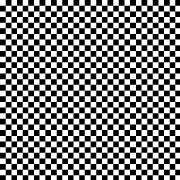}
		\caption{Pixel CB}
		\label{fig:pattern-pixel-cb}
	\end{subfigure}
	
	\caption{Compact visual summary of the eight trigger constructions
		evaluated in the matched CIFAR-10 study. S\&P denotes salt-and-pepper
		noise; CB denotes checkerboard.}
	\label{fig:pattern-overview}
\end{figure}

\paragraph{Compared patterns.}
The first two patterns serve as heuristic controls. Random noise draws RGB-coordinate values from the continuous uniform distribution on $[-1,1]$. Salt-and-pepper noise draws spatial values from $\{-1,+1\}$ and shares them across RGB channels. The next two patterns are the spectral-initialized numerical constructions obtained by empirically estimated RGB (chromatic) precision into Equation~\eqref{eq:ideal-mahalanobis}, first without and then with the luminance restriction in Equation~\eqref{eq:luminance-objective}. Horizontal and vertical period-2 stripes alternate sign between consecutive one-pixel rows and columns, respectively. They are canonical candidates motivated by an eight-neighbor prior whose diagonal-to-axial weighting can change their ranking relative to the pixel checkerboard. Although the two orientations have equal graph energy on a square grid, we evaluate both so that the experiment does not assume orientation-invariant acquisition by the victim. The $2\!\times\!2$ block checkerboard is induced by a four-neighbor lattice over nonoverlapping $2\!\times\!2$ patches. Finally, the pixel checkerboard is the closed-form result of the adopted four-neighbor pixel lattice. Figure~\ref{fig:pattern-overview} provides a compact visual illustration of all eight constructions. Appendix~\ref{app:pattern-constructions} derives the empirical, eight-neighbor, and coarse-patch patterns. We use the same CIFAR-10 configuration as in Section~\ref{sec:attack-performance} to conduct this analysis and results are reported in Table~\ref{tab:pattern-study}.

The pixel checkerboard achieves the highest ASR in Table~\ref{tab:pattern-study}, supporting the practical value of combining the luminance restriction with the four-neighbor prior in this matched setting. By contrast, random noise and salt-and-pepper noise attain only $0.64\%$ and $4.15\%$ ASR, respectively, indicating that arbitrary random patterns are ineffective under the same protocol. The empirical comparison further isolates the effect of the luminance restriction: the unrestricted RGB construction in Equation~\eqref{eq:ideal-mahalanobis} attains $10.23\%$ ASR, whereas the luminance-restricted construction in Equation~\eqref{eq:luminance-objective} attains $93.48\%$, supporting the use of luminance restriction. However, the empirical luminance-restricted construction requires access to all training samples, estimation and inversion of a high-dimensional sample covariance, and numerical optimization. Its estimated pattern in Figure~\ref{fig:pattern-overview}(d) resembles the pixel checkerboard but contains irregular deviations that may reflect instability of the raw sample-covariance estimate or its inversion. In contrast, the four-neighbor prior yields the pixel checkerboard analytically, without empirical data, covariance estimation, or numerical optimization. The $2\!\times\!2$ block checkerboard attains $33.93\%$ ASR, illustrating that trigger acquisition is also sensitive to the imposed spatial scale under this matched configuration. H-stripes and V-stripes attain $58.44\%$ and $84.01\%$ ASR, respectively, both below the pixel checkerboard. More generally, the eight-neighbor prior does not uniquely imply a stripe: among these canonical candidates, the diagonal-to-axial weight $\eta$ determines whether a stripe or the pixel checkerboard ranks higher, with a tie at the threshold. Thus, unlike the four-neighbor construction, its resulting pattern depends on an additional parameter choice.

\begin{table}[t]
	\centering
	\caption{Matched CIFAR-10 trigger-pattern comparison 
		(target 0, $m=20$, $\alpha=10/255$,$\gamma=1$).}
	\label{tab:pattern-study}
	\setlength{\tabcolsep}{4pt}
	\begin{tabular}{lc}
		\toprule
		Pattern & ASR \\
		\midrule
		Random noise $[-1,1]$                         & 0.64\%  \\
		Salt-and-pepper $\{-1,+1\}$                   & 4.15\%  \\
		Empirical Eq.~\eqref{eq:ideal-mahalanobis}    & 10.23\%  \\
		Empirical Eq.~\eqref{eq:luminance-objective}  & 93.48\% \\
		H-stripes (period 2; eight-neighbor)           & 58.44\% \\
		V-stripes (period 2; eight-neighbor)           & 84.01\% \\
		$2\!\times\!2$ block checkerboard              & 33.93\% \\
		Pixel checkerboard (default)                   & 95.72\% \\
		\bottomrule
	\end{tabular}
\end{table}

\FloatBarrier

\section{Defense Evaluation}
\label{sec:defenses}

\begin{table*}[t]
	\centering
	\caption{Summary of 7 evaluated defenses against Checkerboard.}
	\label{tab:defense-summary}
	\small
	\setlength{\tabcolsep}{4pt}
	\begin{tabularx}{\textwidth}{@{}lllcYY@{}}
		\toprule
		Level & Defense & Output & Clean calibration
		& CIFAR-10 outcome & IN-100 outcome \\
		\midrule
		
		\multirow{4}{*}{\makecell[l]{Model\\level}}
		& BTI-DBF
		& Trigger recovery
		& 5\%
		& No dominant target class; recovery failed
		& No dominant target class; recovery failed \\
		
		& MM-BD
		& Model detection
		& --
		& Target score: 0.25; $p=0.99$ (not detected)
		& Target score: 0.86; $p=0.99$ (not detected) \\
		
		& ABL
		& Training-time mitigation
		& --
		& ACC/ASR: 91.37/97.63
		& ACC/ASR: 67.83/84.39 \\
		
		& FT-SAM
		& Model repair
		& 5\%
		& ACC/ASR: 94.54/94.81
		& ACC/ASR: 69.85/83.51 \\
		
		\midrule
		
		\multirow{3}{*}{\makecell[l]{Data\\level}}
		& PoisonSpot
		& Poison detection
		& 5\%
		& TPR/FPR: 5.00/73.84
		& N/A due to resource limit \\
		
		& Beatrix
		& Poison detection
		& 30 samples per class
		& Score: 1.30 (not detected)
		& Score: 0.53 (not detected) \\
		
		& FREAK
		& Test-input detection
		& 160 samples
		& TPR/FPR: 49.05/13.86
		& TPR/FPR: 30.08/6.79 \\
		
		\bottomrule
	\end{tabularx}
\end{table*}

We evaluate Checkerboard against 7 representative backdoor defenses (with an additional 12 defenses reported in Appendix~\ref{app:additional-defenses}), covering model- and data-level approaches, as well as adaptive defenses. \emph{Model-level} defenses inspect an infected model, modify its training, or repair the trained parameters; \emph{data-level} defenses decide whether training samples, or deployment inputs are suspicious.  Following prior work~\cite{zeng2023narcissus, yu2024generalization}, we use CIFAR-10 as the primary benchmark because it is supported by most existing defenses, enabling consistent and comprehensive comparison. We further evaluate on IN-100 to test whether Checkerboard’s defense resistance extends to a higher-resolution, more complex setting. The Checkerboard setting remains unchanged from Section~\ref{sec:attack-performance}.

\subsection{Model-Level Defenses}
\label{sec:model-level-defenses}
\begin{figure}[t]
	\centering
	\begin{subfigure}{0.32\columnwidth}
		\centering
		\includegraphics[width=\linewidth]{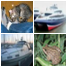}
		\caption{Clean images}
	\end{subfigure}
	\begin{subfigure}{0.32\columnwidth}
		\centering
		\includegraphics[width=\linewidth]{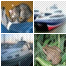}
		\caption{Poisoned images}
	\end{subfigure}
	\begin{subfigure}{0.32\columnwidth}
		\centering
		\includegraphics[width=\linewidth]{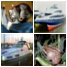}
		\caption{Recovered images}
	\end{subfigure}
	
	\caption{Visual analysis using BTI-DBF on a Checkerboard-poisoned CIFAR-10 model shows that recovered samples remain indistinguishable from clean images, demonstrating a failure to accurately synthesize the Checkerboard trigger.}
	\label{fig:bti-dbf}
\end{figure}

\textbf{BTI-DBF} (ICLR 24) \cite{xu2024towards} synthesizes candidate triggers in feature space by learning a generator that separates clean content from potential trigger features. Following the original paper, we evaluate BTI-DBF using 5\% clean samples on both CIFAR-10 and IN-100. Under a successful recovery, the synthesized pattern should induce a dominant target prediction. Instead, the generated samples from BTI-DFB produce an approximately uniform class distribution on both datasets. This indicates that the recovered pattern does not activate any consistent target behavior, and BTI-DBF fails to synthesize the checkboard trigger. One plausible explanation is that BTI-DBF relies on spatial decoupling, whereas Checkerboard is a dense global perturbation that overlaps with clean content across the full image. Therefore, it fails to isolate the backdoor signal, consistent with the examples in Figure~\ref{fig:bti-dbf}.

\begin{figure}[t]
	\centering
	\includegraphics[width=\linewidth]{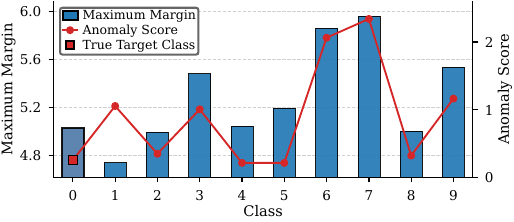}
	\caption{Statistics from MM-BD on CIFAR-10.}
	
	\label{fig:mmbd}
\end{figure}

\textbf{MM-BD} (SP 24) \cite{wang2024mm} is a post-training backdoor detection method that identifies the target class by testing whether its maximum-margin statistic forms a significant outlier among all classes. Specifically, MM-BD fits a null distribution to the maximum-margin statistics of the non-maximum classes and declares a model backdoored when the resulting order-statistic $p$-value is below the default threshold of $0.05$. When applying MM-BD to Checkerboard-infected models, Checkerboard does not induce the expected distinctive maximum-margin pattern. On CIFAR-10, as shown in Figure~\ref{fig:mmbd}, the maximum margins of all classes are concentrated in a narrow range, and the true target class has a maximum margin of only 5.03, which lies within the range of the other classes (4.73--5.96) rather than standing clearly above them. Its MAD-based anomaly score is also only 0.25. Correspondingly, MM-BD yields a $p$-value of 0.99, which is far above the detection threshold of $0.05$, and therefore does not flag the model as backdoored. Similarly, on IN-100, the true target class has an anomaly score of only 0.86, and MM-BD also produces a $p$-value of 0.99. Thus, on both datasets, the true target class does not exhibit the statistically significant maximum-margin outlier required by MM-BD, causing it to fail to detect the attack.

\begin{figure}[t]
	\centering
	\includegraphics[width=\linewidth]{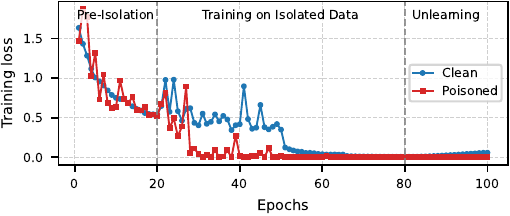}
	\caption{Loss values of ABL in three stages on CIFAR-10.}
	\label{fig:ABL}
\end{figure}

\textbf{ABL} (NeurIPS 21) \cite{li2021anti} assumes that poisoned samples converge faster than clean samples and therefore appear as unusually low-loss examples early in training. It identifies these samples, excludes them from subsequent training, and then performs unlearning using the identified samples to unlearn backdoor. On Checkerboard-poisoned data, ABL reduces ACC from 94.24\% to 91.37\% while leaving the ASR at 97.63\% on CIFAR-10; on IN-100, it reduces ACC from 69.94\% to 67.83\% while retaining an ASR of 84.39\% at $\gamma=1$. Thus, the defense fails both to preserve clean accuracy and to remove the backdoor. Figure~\ref{fig:ABL} shows the ABL's CIFAR-10 loss value evolution, the poisoned samples do not form the expected low-loss group; their losses remain comparable to, or even higher than, those of clean samples, with a similar trend observed on IN-100. This prevents ABL from isolating the true poisons and instead causes it to discard clean data.

\textbf{FT-SAM} (ICCV 23) \cite{zhu2023enhancing} combines sharpness-aware minimization with targeted penalties on large-norm neurons, based on the assumption that backdoor features are encoded by a small set of dominant parameters.  We perform FA-SAM to fine-tune Checkerboard-infected model on CIFAR-10 and IN-100 using 5\% clean data. On CIFAR-10, FT-SAM yields 94.54\% ACC and 94.81\% ASR; On IN-100, it achieves 69.85\% ACC while leaving the ASR at 83.51\% under $\gamma=1$, indicating negligible mitigation. A plausible explanation is that the Checkerboard signal is not concentrated in the heavily penalized large-norm neurons, and instead remains distributed across features that FT-SAM does not aggressively suppress.

\subsection{Data-Level Defenses}
\label{sec:data-level-defenses}

\begin{figure}[t]
	\centering
	\includegraphics[width=\linewidth]{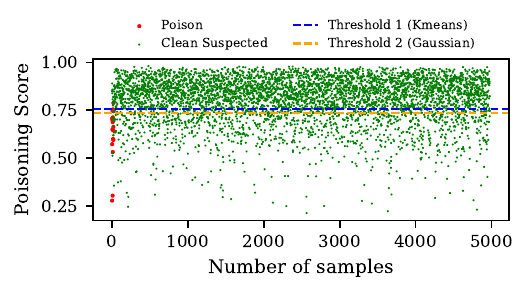}
	\caption{PoisonSpot's poisoning score on CIFAR-10.}
	
	\label{fig:PoisonSpot}
\end{figure}

\textbf{PoisonSpot} (CCS 25)~\cite{10.1145/3719027.3744802} is specifically designed to detect clean-label backdoor attacks through class-wise training provenance tracking, monitoring how individual samples influence sensitive model parameters. It computes poisoning scores by comparing the gradient trajectories of suspected samples against a clean baseline, where scores near 1.0 indicate anomalous behavior and scores near 0.0 indicate benign samples. PoisonSpot further uses two dynamic decision boundaries: Threshold 1, derived from K-means clustering, and Threshold 2, estimated using a Gaussian Mixture Model. We evaluate PoisonSpot against Checkerboard on CIFAR-10 using 5\% clean samples as the baseline. Notably, even on CIFAR-10, its peak memory usage exceeds 100~GB in our experiments; therefore, we are unable to scale the defense to IN-100 under our available computational resources. On CIFAR-10, PoisonSpot exhibits a severe 73.84\% FPR while achieving only a 5\% TPR. This failure arises because Checkerboard violates PoisonSpot's assumption of separability between clean and poisoned samples. As shown in Figure~\ref{fig:PoisonSpot}, clean samples exhibit substantial variance in poisoning scores, forming a dense band between 0.4 and 0.95 that overwhelms the 20 poisoned samples. Consequently, PoisonSpot fails to distinguish Checkerboard poisons from inherent training dynamics, causing both decision boundaries to intersect the clean-sample distribution and resulting in excessive false positives while leaving most poisoned samples undetected.

\begin{figure}[t]
	\centering
		
	\begin{subfigure}{0.49\columnwidth}
		\centering
		\includegraphics[width=\linewidth]{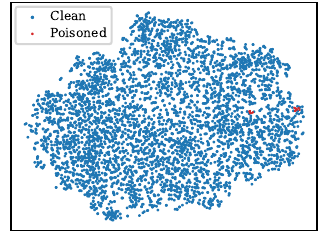}
		\caption{CIFAR-10}
	\end{subfigure}
	\hfill 
	\begin{subfigure}{0.49\columnwidth}
		\centering
		\includegraphics[width=\linewidth]{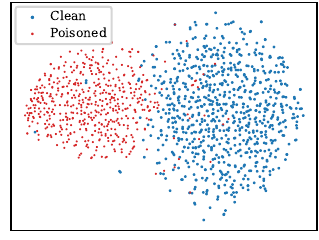}
		\caption{IN-100}
	\end{subfigure}

	\caption{Visualization of Gramian features of Beatrix.}
	\label{fig:beatrix}
	\vspace*{-10pt}
\end{figure}

\textbf{Beatrix} (NDSS 23)~\cite{ma2023beatrix} detects backdoors by using Gram matrices to characterize high-order feature correlations and regularized maximum mean discrepancy to identify anomalous class mixtures. Following the original setup, we use 30 clean samples per class to calibrate the detector on both CIFAR-10 and IN-100. Nevertheless, Beatrix fails to detect Checkerboard: the anomaly scores of the target class are only 1.30 on CIFAR-10 and 0.53 on IN-100, both well below the detection threshold of $e^2\approx 7.39$. Figure~\ref{fig:beatrix} visualizes Gramian features extracted from the AveragePooling layer of ResNet-18 and projected using t-SNE. On CIFAR-10, poisoned samples do not form the distinct secondary cluster assumed by Beatrix, but instead largely overlap with clean target-class samples. On IN-100, although clean and poisoned samples occupy somewhat different regions in the t-SNE projection, they still do not exhibit a clear, statistically detectable mixture structure, consistent with the low anomaly score.

\begin{figure}[t]
	\centering
	\includegraphics[width=\linewidth]{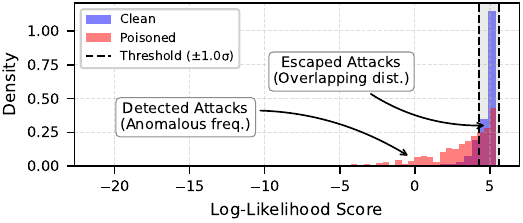}
	\caption{Log-Likelihood dist.  in FREAK on CIFAR-10.}
	
	\label{fig:freak}
\end{figure}

\textbf{FREAK} (CVPR 23) \cite{al2023don} is a test-time detector specifically designed for high-frequency backdoors. It analyzes the sensitivity of the model’s top logit to Fourier-magnitude perturbations and flags anomalies using a calibrated log-likelihood threshold. Using 160 clean samples for calibration, FREAK achieves 49.05\% TPR and 13.86\% FPR on CIFAR-10, and 30.08\% TPR and 6.79\% FPR on IN-100. Figure~\ref{fig:freak} shows that the log-likelihood distributions of clean and poisoned samples overlap substantially on CIFAR-10. This indicates that although Checkerboard is high-frequency in structure, its low-amplitude blended form does not produce the strong frequency anomaly that FREAK expects.

\subsection{Adaptive Defenses}
\label{sec:adaptive-defenses}

We further evaluate the adaptive defenses designed with knowledge of Checkerboard's structure. 

\paragraph{Checkerboard-Notch.}
We first consider a strong adaptive defender that knows the spatial pattern of the trigger. Checkerboard-Notch estimates the checkerboard component of each input and subtracts it before classification. We use the strongest evaluated configuration, which removes the full estimated projection. Implementation details and sanitized examples are provided in Appendix~\ref{sec_appendix_checkerboard_notch}.

Using the same attack and victim-model settings as in Section~\ref{sec:attack-performance}, Checkerboard-Notch retains ACC of 94.72\% on CIFAR-10 and 69.12\% on IN-100. At the native test perturbation, it reduces the ASR to 50.04\% and 53.30\%, respectively. However, when the test trigger is amplified twice, the ASR increases to 89.19\% on CIFAR-10 and 84.58\% on IN-100. Thus, explicitly removing the estimated checkerboard component substantially weakens the attack at its native test strength, but does not eliminate it under stronger test-time amplificaation.

\paragraph{Gaussian blur.}
We next evaluate Gaussian blur as a low-pass preprocessing defense. The default Checkerboard trigger alternates at the pixel level, it contains strong high-frequency components and is a natural target for blur-based sanitization. We apply Gaussian blur with $\sigma\in\{0.5,1.0,2.0\}$ using a $3\times3$ kernel. The complete results are reported in Table~\ref{tab:adaptive-defense} in Appendix~\ref{sec_appendix_preprocessing} and summarized in this section.

We find weak blur with $\sigma=0.5$ retains native ASRs of 95.97\% on CIFAR-10 and 80.26\% on IN-100. In contrast, strong blur with $\sigma=1.0$ and $2.0$ reduce the native ASR to near 0 on CIFAR-10 and 9.68--12.53\% on IN-100, showing low-pass blurring can suppress the default checkerboard trigger.

To adapt to this defense, we lower the spatial frequency of the checkerboard trigger by using the $2\times2$ block pattern introduced in Section~\ref{sec:pattern-study} and illustrated in Figure~\ref{fig:pattern-overview}. Its sign changes every two pixels rather than every pixel, making it more resistant to low-pass filtering. We also increase the poison count from 20 to 100 on CIFAR-10 and from 520 to 650 on IN-100, while retaining the respective training perturbations of $\alpha=10/255$ and $16/255$. Thus, this experiment jointly adapts the trigger frequency and poison budget rather than isolating the effect of block size.

Under $\sigma=1.0$ and $2.0$, the adaptive configuration achieves native ASRs of 51.31--70.84\% on CIFAR-10 and 53.25--55.63\% on IN-100. With stronger test-time amplification, ASR increase to 98.42--99.84\% at $\times2$ on CIFAR-10 and 83.82--84.39\% at $\times1.5$ on IN-100. These results indicate that Gaussian blur effectively targets the default high-frequency instantiation, while a lower-frequency configuration can recover substantial attack effectiveness by jointly adjusting the trigger scale, poison count, and test-time activation. Appendix~\ref{sec_appendix_preprocessing} additionally reports results for mean filtering, JPEG compression, and DCT-based high-frequency suppression.

\section{Discussion \& Conclusion}


We present Checkerboard, a learning-free clean-label backdoor attack with closed-form, data-independent trigger design. Starting from an input-space Fisher-separability proxy, we derive the pixel-wise checkerboard as a global maximizer under the stated luminance restriction and ridge four-neighbor natural-image prior. Trigger construction requires no surrogate model, data access or optimization.

Experiments on four benchmark datasets show that Checkerboard outperforms nine baseline attacks under low poisoning budgets. Because the trigger is generated analytically rather than learned or optimized, Checkerboard incurs negligible poison-generation cost. These results suggest that the design of clean-label backdoors can be much simpler than existing learning-based attacks: even a fixed input-space pattern can induce strong backdoor behavior.

Our defense evaluations further show that Checkerboard remains resilient under representative and adaptive defenses. Although the default pixel-wise trigger is sensitive to strong low-pass preprocessing due to its high-frequency structure, this limitation does not fundamentally weaken the attack. Pattern-aware sanitization attenuates but does not eliminate the learned backdoor behavior, and preprocessing defenses can be resisted by increasing the checkerboard side length. While this lower-frequency adaptation may require moderately more poisoned samples, the perturbation magnitude remains unchanged and the overall poisoning budget stays low. Thus, the observed limitation mainly reflects an attacker-side tradeoff between separability and preprocessing robustness, rather than a fundamental defense against Checkerboard.


\phantomsection\label{audit:body-end}

\bibliographystyle{plain}
\bibliography{sample-base}

\appendix
\section{FDR Derivations}
\label{app:theory}

\subsection{Maximum FDR for Two Distributions}

Let distributions $P_0$ and $P_1$ have means $\bmu_0,\bmu_1$ and covariances $\bSigma_0,\bSigma_1$, and assume $S=\bSigma_0+\bSigma_1\succ0$. For a nonzero projection $\bm{w}$, their Fisher discriminant ratio is
\begin{equation}
  \mathcal{J}(\bm{w})=
  \frac{\bigl(\bm{w}^{\top}(\bmu_1-\bmu_0)\bigr)^2}
       {\bm{w}^{\top}S\bm{w}}.
\end{equation}
Writing $\bm{v}=S^{1/2}\bm{w}$ and $\bm{u}=S^{-1/2}(\bmu_1-\bmu_0)$ gives
\(
\mathcal{J}(\bm{w})=(\bm{v}^{\top}\bm{u})^2/\|\bm{v}\|_2^2
\).
Cauchy--Schwarz therefore yields
\begin{equation}
  \max_{\bm{w}\ne0}\mathcal{J}(\bm{w})
  =(\bmu_1-\bmu_0)^\top S^{-1}(\bmu_1-\bmu_0),
  \label{eq:appendix-general-fdr}
\end{equation}
When $\bmu_1\ne\bmu_0$, a maximizing direction is proportional to $S^{-1}(\bmu_1-\bmu_0)$. If $\bmu_1=\bmu_0$, the maximum FDR is zero for every nonzero projection.

\subsection{Unclipped Additive Trigger}

Let $\bx\sim\mathcal{P}_x$ have mean $\bmu_x$ and covariance $\bSigma_x\succ0$, and let $\bp=\bx+\alpha\bdelta$. Then
\begin{equation}
  \bmu_p-\bmu_x=\alpha\bdelta,
  \qquad
  \bSigma_p=\bSigma_x.
\end{equation}
As in Section~\ref{sec:separability-objective}, $\mathcal{P}_p^{\bdelta,\alpha}$ is the distribution induced by applying the same fixed candidate trigger to every draw from $\mathcal{P}_x$; it is not the finite mixture used for victim training.
Substituting these quantities into Equation~\eqref{eq:appendix-general-fdr} gives
\begin{equation}
  \operatorname{Sep}_{\mathrm{FDR}}
  (\mathcal{P}_x,\mathcal{P}_p^{\bdelta,\alpha})
  =\alpha^2\bdelta^\top(2\bSigma_x)^{-1}\bdelta
  =\frac{\alpha^2}{2}\bdelta^\top\bSigma_x^{-1}\bdelta,
\end{equation}
which proves Equation~\eqref{eq:fdr-unclipped}. For $\bdelta\ne\bm{0}$, the maximizing projection is $\bm{w}^{*}\propto\bSigma_x^{-1}\bdelta$; for $\bdelta=\bm{0}$, the FDR is zero. Hence, for fixed $\alpha$, the ordering of candidate triggers is exactly the ordering induced by their Mahalanobis magnitude $\bdelta^\top\bSigma_x^{-1}\bdelta$.

\section{Checkerboard Pattern under Clipping}
\label{app:clipping}

This section demonstrates whether enforcing the valid pixel range changes the spatial
pattern selected by the analysis. We first state the exact, distribution-dependent
clipped FDR and then show that, under a common coordinatewise clipping response
and the four-neighbor contrast model used in the main derivation, the selected
pattern remains checkerboard.

\subsection{Exact Distribution-Dependent Formulation}

Let $\bm{X}\sim\mathcal{P}_x$ have covariance $\bSigma_x$ and, for a candidate luminance pattern $\bg$, define
\begin{equation}
  \bm{P}_{\bg}=T_{\bg,\alpha}(\bm{X})
  =\operatorname{clip}(\bm{X}+\alpha R\bg,0,1),
  \qquad r_{\bg}(\bm{X})=\bm{P}_{\bg}-\bm{X}.
\end{equation}
Denote
\begin{equation}
  \bm{m}_{\bg}=\mathbb{E}[r_{\bg}(\bm{X})],
  \qquad \bSigma_{p,\bg}=\operatorname{Cov}(\bm{P}_{\bg}).
\end{equation}
The mean displacement from the clean distribution to its clipped-triggered counterpart is $\bm{m}_{\bg}$. Consequently, their exact maximum FDR is
\begin{equation}
  J_{\mathrm{clip}}(\bg)
  =\bm{m}_{\bg}^{\top}
  (\bSigma_x+\bSigma_{p,\bg})^{-1}\bm{m}_{\bg}.
  \label{eq:exact-clipped-fdr}
\end{equation}
Both $\bm{m}_{\bg}$ and $\bSigma_{p,\bg}$ depend on the image distribution and candidate pattern. The exact clipped problem is therefore distribution-dependent. To come up with a data-independent construction, we introduce a common coordinatewise response together with the same four-neighbor contrast model used in the main derivation.

\subsection{Clipping-Aware Local-Smoothness Model}

For a scalar coordinate $Z\in[0,1]$ and $s\in[-1,1]$, define its mean clipping response
\begin{equation}
  \psi_{\alpha}(s)
  =\mathbb{E}\!\left[\operatorname{clip}(Z+\alpha s,0,1)-Z\right].
  \label{eq:clipping-response}
\end{equation}
For every fixed $Z$, the integrand is nondecreasing in $s$, so $\psi_\alpha$ is nondecreasing. 

For the clipping-aware surrogate, assume that the replicated scalar coordinates have a common marginal response, so the same function $\psi_\alpha$ applies across spatial locations and channels. Define $\bm{\psi}_\alpha(\bg)=(\psi_\alpha(g_1),\ldots,\psi_\alpha(g_n))^\top$. The effective RGB mean displacement is then $\bm{m}_{\bg}=R\bm{\psi}_\alpha(\bg)$. Denote
\begin{equation*}
  \bQ_{\mathrm{clip},\bg}
  :=R^\top(\bSigma_x+\bSigma_{p,\bg})^{-1}R,
\end{equation*}
as the exact clipped FDR in Equation~\eqref{eq:exact-clipped-fdr} equals
$\bm{\psi}_\alpha(\bg)^\top\bQ_{\mathrm{clip},\bg}
\bm{\psi}_\alpha(\bg)$ under this common-response assumption.
We then adopt the clipping-aware counterpart of the main local-smoothness model: the contrast component of $\bQ_{\mathrm{clip},\bg}$ is approximated by a fixed positively weighted four-neighbor Laplacian whose graph and edge weights do not depend on the sign arrangement of $\bg$. The resulting contrast objective is, up to a positive scale,
\begin{equation}
  \widetilde{J}_{\mathrm{clip}}(\bg)
  =\sum_{(i,j)\in E}w_{ij}
  \left[\psi_\alpha(g_i)-\psi_\alpha(g_j)\right]^2.
  \label{eq:clipped-laplacian-surrogate}
\end{equation}
The ridge $\beta I$ in A1 is introduced to make the unclipped precision surrogate positive definite. After clipping, its contribution would be proportional to $\|\bm{\psi}_\alpha(\bg)\|_2^2$, which need not be constant without additional response symmetry. Equation~\eqref{eq:clipped-laplacian-surrogate} and the result below therefore concern the four-neighbor contrast component explicitly stated in Section~\ref{sec:attack-construction}.

Because $\psi_\alpha$ is nondecreasing, for any $g_i,g_j\in[-1,1]$,
\begin{equation}
  \left|\psi_\alpha(g_i)-\psi_\alpha(g_j)\right|
  \le \psi_\alpha(1)-\psi_\alpha(-1).
\end{equation}
Thus, the largest possible response contrast on an edge is attained by assigning one endpoint $+1$ and the other $-1$. The four-neighbor grid is bipartite, so assigning $+1$ to one partition and $-1$ to the other reaches this upper bound on every positively weighted edge simultaneously. The global sign flip produces the same squared edge differences. Consequently, under the common clipping-response model and the fixed positively weighted four-neighbor local-smoothness model, the two pixel-wise checkerboard phases are global maximizers of the clipping-aware contrast objective in Equation~\eqref{eq:clipped-laplacian-surrogate}.

Clipping therefore replaces the attainable adjacent-pixel response contrast by $\psi_\alpha(1)-\psi_\alpha(-1)$, but it does not change the maximizing alternating sign arrangement. Hence, under the adopted clipping-aware common-response and four-neighbor contrast model, the resulting trigger pattern is still the pixel-wise checkerboard, up to a global sign flip. This conclusion does not require clipping to be rare or the pixel distribution to be reflection-symmetric. It is nevertheless conditional on the stated approximation: if clipping makes the induced pooled precision depend materially on the candidate sign arrangement or the marginal response varies strongly across coordinates, the exact optimizer in Equation~\eqref{eq:exact-clipped-fdr} can be distribution-dependent.

\section{Comparator Constructions and Alternative Priors}
\label{app:graph-sensitivity}

The four-neighbor structure adopted in Section~\ref{sec:graph-prior} is a natural-image approximation rather than an estimate of the exact covariance of a particular dataset. Here we first derive its checkerboard solution and then examine what changes when that structure is replaced. We distinguish three construction requirements. \emph{Attack-model-free} means that the prior uses no victim, surrogate, generator, or pretrained attack-side model. \emph{Data-free} means that no image samples, including target-class images later used as poison carriers, are used to estimate the graph or its weights. \emph{Parameter-free} means that no fitted or tuned numerical graph coefficient affects the maximizing pattern; selecting a graph topology remains an explicit modeling assumption.

\subsection{Solving the Adopted Four-Neighbor Objective}
\label{app:four-neighbor-solution}

For every feasible $\bg$ in Equation~\eqref{eq:graph-objective}, $(g_i-g_j)^2\leq4$ on each edge and $g_i^2\leq1$ at each node. Consequently,
\begin{equation*}
  \widetilde{a}(\bg)
  \leq 4\lambda\sum_{(i,j)\in E}w_{ij}+\beta n
  =:U.
\end{equation*}
The ordinary four-neighbor grid is bipartite. Let $V_+=\{(u,v):u+v\text{ is even}\}$ and $V_-=\{(u,v):u+v\text{ is odd}\}$. Every edge joins the two partitions, so assigning $+1$ to $V_+$ and $-1$ to $V_-$ makes every node attain $g_i^2=1$ and every edge attain $(g_i-g_j)^2=4$. The resulting objective value is exactly $U$; reversing all signs gives the same value. The two checkerboard phases in Equation~\eqref{eq:checkerboard} therefore solve Equation~\eqref{eq:graph-objective}.

They are also the only maximizers. If $\widehat{\bg}$ attains $U$, then
\begin{align*}
  0=U-\widetilde{a}(\widehat{\bg})
  &=\lambda\sum_{(i,j)\in E}w_{ij}
    \left[4-(\widehat g_i-\widehat g_j)^2\right]\\
  &\quad+\beta\sum_{i\in V}\left(1-\widehat g_i^2\right).
\end{align*}
Every bracketed quantity is nonnegative. Since $\lambda>0$, $\beta>0$, and every $w_{ij}>0$, equality forces $|\widehat g_i|=1$ at every node and opposite signs across every edge. Connectedness propagates either initial sign throughout the grid, leaving only the two global phases. Thus $\lambda$, $\beta$, and the positive edge weights determine the optimum value $U$, but not the maximizing pattern; this is why they do not appear in Equation~\eqref{eq:checkerboard}.

\subsection{General Graph Consequence}

Let $G=(V,E)$ be any fixed positively weighted graph and consider
\begin{equation}
  F_G(\bg)=\lambda\bg^\top L_G\bg+\beta\|\bg\|_2^2,
  \qquad \bg\in[-1,1]^n,
  \label{eq:general-graph-prior}
\end{equation}
with $\lambda>0$ and $\beta>0$. Because this is a convex quadratic maximized over a hypercube, at least one maximizer lies at a vertex $\bm{s}\in\{-1,+1\}^n$. For such a binary pattern,
\begin{equation}
  F_G(\bm{s})
  =4\lambda\!\sum_{(i,j)\in\operatorname{Cut}(\bm{s})}\!w_{ij}
   +\beta n.
  \label{eq:graph-max-cut}
\end{equation}
Thus a positive graph-Laplacian prior maps directly to a maximum-weight-cut pattern. If $G$ is connected and bipartite, its two bipartition assignments cut every edge and are the two maximizers. If $G$ is non-bipartite, not all local contrast preferences can generally be satisfied simultaneously, and the solution depends on the topology and relative weights.

\subsection{Derivation of the Comparison Patterns}
\label{app:pattern-constructions}

Section~\ref{sec:pattern-study} compares two random controls, five patterns derived here, and the pixel checkerboard. Random noise and salt-and-pepper noise are operational controls rather than solutions of a precision model, so Section~\ref{sec:pattern-study} defines them directly. The pixel checkerboard is already derived from the adopted four-neighbor prior in Appendix~\ref{app:four-neighbor-solution}. This subsection derives the two empirical-precision patterns, the horizontal and vertical stripe candidates associated with an eight-neighbor prior, and the $2\!\times\!2$ block checkerboard associated with a coarse patch lattice.

\paragraph{Empirical RGB precision: Equation~\eqref{eq:ideal-mahalanobis}.}
Let $N$ and let $\{\bx_k\}_{k=1}^{N}$ be the clean CIFAR-10 training images flattened in RGB space, with $d$ the number of pixels per image ($N=50000$ and $d=3072$ on CIFAR-10). From the empirical mean $\bar{\bx}$, we form the ordinary unbiased sample covariance
\begin{equation}
  S
  :=\frac{1}{N-1}\sum_{k=1}^{N}
  (\bx_k-\bar{\bx})(\bx_k-\bar{\bx})^\top,
  \qquad
  \widehat{\bm{\Omega}}_{\rm raw}:=S^{-1},
  \label{eq:empirical-rgb-precision}
\end{equation}
where the definition of $\widehat{\bm{\Omega}}_{\rm raw}$ assumes that $S$ is nonsingular. Substituting this raw-sample precision into Equation~\eqref{eq:ideal-mahalanobis} gives
\begin{equation}
  \max_{\bdelta\in[-1,1]^d}
  \bdelta^\top\widehat{\bm{\Omega}}_{\rm raw}\bdelta.
  \label{eq:empirical-eq5}
\end{equation}
Because $\widehat{\bm{\Omega}}_{\rm raw}$ is positive definite, at least one maximizer lies at a binary vertex, but globally maximizing the resulting binary quadratic form is generally intractable. We therefore use a reproducible spectral initialization. Let $\bm v_5$ be a unit leading eigenvector of $\widehat{\bm{\Omega}}_{\rm raw}$, equivalently an eigenvector of $S$ associated with its smallest eigenvalue. It exactly solves the Euclidean relaxation
\begin{equation}
  \max_{\|\bdelta\|_2^2\le d}
  \bdelta^\top\widehat{\bm{\Omega}}_{\rm raw}\bdelta
  =d\lambda_{\max}(\widehat{\bm{\Omega}}_{\rm raw}),
  \qquad
  \bdelta_{\rm rel}=\sqrt d\,\bm v_5.
  \label{eq:empirical-eq5-relaxation}
\end{equation}
The relaxed vector need not satisfy the coordinatewise bound. We initialize a feasible full-amplitude pattern by $\bm s_5^{(0)}=\operatorname{sign}(\bm v_5)$, assigning $+1$ to any zero entry, and then perform monotone one-bit refinement: a coordinate is flipped only when the flip strictly increases the objective in Equation~\eqref{eq:empirical-eq5}. The process terminates at a one-flip local optimum, which we denote by $\widehat{\bdelta}_5$. It is a spectral-initialized numerical solution, not a certificate of the global box-constrained optimum.

\paragraph{Empirical luminance-restricted precision: Equation~\eqref{eq:luminance-objective}.}
To isolate the luminance-restricted in Equation~\eqref{eq:luminance-objective}, we reuse the same raw-sample RGB precision $\widehat{\bm{\Omega}}_{\rm raw}$ rather than estimating a separate grayscale covariance. The induced empirical precision is
\begin{equation}
  \widehat{\bQ}_{\rm lum,raw}
  :=R^\top\widehat{\bm{\Omega}}_{\rm raw}R,
  \qquad
  \max_{\bg\in[-1,1]^n}
  \bg^\top\widehat{\bQ}_{\rm lum,raw}\bg.
  \label{eq:empirical-eq6}
\end{equation}
Let $\bm v_6$ be a unit leading eigenvector of $\widehat{\bQ}_{\rm lum,raw}$. We initialize $\bm s_6^{(0)}=\operatorname{sign}(\bm v_6)$ and apply the same strictly improving one-bit refinement under Equation~\eqref{eq:empirical-eq6}. The resulting spatial pattern $\widehat{\bg}_6$ is replicated as $\widehat{\bdelta}_6=R\widehat{\bg}_6$. Consequently, the comparison between $\widehat{\bdelta}_5$ and $\widehat{\bdelta}_6$ changes only the feasible set: the former may vary independently across RGB coordinates, whereas the latter is channel-shared and luminance-restricted. 

\paragraph{Eight-neighbor prior: checkerboard or stripes.}
Let $E_{\rm a}$ contain horizontal and vertical edges and let $E_{\rm d}$ contain both diagonal edge families. With unit axial weight and diagonal-to-axial ratio $\eta>0$, the eight-neighbor objective is
\begin{equation}
  F_{8,\eta}(\bg)
  =\lambda\!\sum_{(i,j)\in E_{\rm a}}(g_i-g_j)^2
   +\lambda\eta\!\sum_{(i,j)\in E_{\rm d}}(g_i-g_j)^2
   +\beta\|\bg\|_2^2.
  \label{eq:eight-neighbor-objective}
\end{equation}
Consider the pixel checkerboard $g^{\rm C}_{u,v}=(-1)^{u+v}$, horizontal stripes $g^{\rm H}_{u,v}=(-1)^u$, and vertical stripes $g^{\rm V}_{u,v}=(-1)^v$. Thus the horizontal pattern alternates between consecutive one-pixel rows, the vertical pattern alternates between consecutive one-pixel columns, and each repeats every two pixels along its varying axis. On the ordinary non-wrapping $H\times W$ grid, direct edge counting gives
\begin{align}
  F_{8,\eta}(\bg^{\rm C})
  &=4\lambda\bigl[(H-1)W+H(W-1)\bigr]+\beta n,\nonumber\\
  F_{8,\eta}(\bg^{\rm H})
  &=4\lambda\bigl[(H-1)W
      +2\eta(H-1)(W-1)\bigr]+\beta n,\nonumber\\
  F_{8,\eta}(\bg^{\rm V})
  &=4\lambda\bigl[H(W-1)
      +2\eta(H-1)(W-1)\bigr]+\beta n.
  \label{eq:eight-neighbor-candidates}
\end{align}
For a square $H=W$, the two stripe orientations tie. Each exceeds the checkerboard candidate when
\begin{equation*}
  \eta>\frac{H}{2(H-1)},
\end{equation*}
whose large-grid limit is $1/2$; below this threshold the checkerboard candidate has the larger value, and equality gives a tie. Equivalently, ignoring boundary terms yields the familiar per-image leading terms $8\lambda n+\beta n$ for the checkerboard and $(4+8\eta)\lambda n+\beta n$ for either stripe. These calculations show how the eight-neighbor objective can rank the canonical checkerboard and stripe candidates differently as $\eta$ varies. They do not freeze $\eta$, establish that a stripe is the unique global maximizer, or exclude boundary-sensitive or tied maximum-cut solutions under a general weighting.

The additional choice is the relative diagonal weight $\eta$. For full-amplitude binary patterns, the ridge contribution is always $\beta n$, and $\lambda>0$ only rescales the graph-energy difference; neither selects checkerboard versus stripes. The trigger magnitude $\alpha$ is likewise not a graph-selection parameter. Thus, unlike the adopted four-neighbor construction, an eight-neighbor prior requires a relative-weight specification that can change the resulting pattern. The two stripe orientations tie in Equation~\eqref{eq:eight-neighbor-candidates} for square images, but Section~\ref{sec:pattern-study} retains both as empirical results so that it does not assume orientation-invariant acquisition. Neither orientation nor $\eta$ is selected using attack success.

\paragraph{Coarse $2\!\times\!2$ patch lattice.}
For even $H$ and $W$, partition the image into nonoverlapping $2\!\times\!2$ patches and constrain every pixel in a patch to share one value:
\begin{equation*}
  g_{u,v}=h_{\lfloor u/2\rfloor,\lfloor v/2\rfloor}.
\end{equation*}
Place the patch variables $h_{p,q}$ on an ordinary four-neighbor grid of size $(H/2)\times(W/2)$. This coarse graph is connected and bipartite, so the argument in Appendix~\ref{app:four-neighbor-solution} gives the two patch-level maximizers $h_{p,q}=\pm(-1)^{p+q}$. Lifting either phase back to pixels yields
\begin{equation}
  g^{\rm B}_{u,v}
  =\pm(-1)^{\lfloor u/2\rfloor+\lfloor v/2\rfloor},
  \label{eq:block-checkerboard}
\end{equation}
the $2\!\times\!2$ block checkerboard. This prior remains model- and data-free, but it introduces a predefined patch scale. In Section~\ref{sec:pattern-study}, it is evaluated at the same poison count and native trigger strength as every other pattern. The separate adaptive-defense experiment uses a different poison budget and is not part of this matched comparison.

In summary, the empirical Equation~\eqref{eq:empirical-eq5} and Equation~\eqref{eq:empirical-eq6} patterns test the cost of replacing an estimated precision by a structural prior, but require data and numerical optimization. The eight-neighbor candidates test sensitivity to diagonal adjacency, but depend on the additional relative weight $\eta$. The coarse-patch construction tests spatial scale, but predefines a patch size. By contrast, the adopted positive four-neighbor prior yields the pixel checkerboard for every positive choice of its numerical coefficients and edge weights.

\section{Detailed Experimental Protocol}
\label{app:experimental-details}

\subsection{Datasets and Label Construction}

CIFAR-10 and CIFAR-100 use their standard training and test splits~\cite{krizhevsky2009learning}. Following prior clean-label backdoor settings~\cite{nguyen2021wanet,huynh2024combat}, CelebA-8 is formed from three relatively balanced CelebA attributes~\cite{liu2015deep}. Let $a_1$, $a_2$, and $a_3$ denote the binary indicators for \emph{Heavy Makeup}, \emph{Mouth Slightly Open}, and \emph{Smiling}, respectively. We assign each image the class
\[
  y = 4a_1 + 2a_2 + a_3,
  \qquad a_1,a_2,a_3\in\{0,1\},
\]
so that the eight classes correspond to the eight possible attribute combinations. CelebA-8 retains the original CelebA split membership. The evaluation uses a fixed IN-100, which is constructed from the full ImageNet-1K. This subset selects the first 100 classes sorted by alphabetically from ImageNet-1k. ImageNet-1k can be obtained from \url{https://huggingface.co/datasets/ILSVRC/imagenet-1k} The training-set sizes, target-0 counts, poison counts, and target-class poison rates are reported in Table~\ref{tab:dataset-budget}.

\subsection{Poison Construction and Evaluation Regimes}

After the trigger is fixed, the adversary uniformly samples $m$ distinct indices from the complete target-class subset $\mathcal{D}_t$. Each selected image is replaced by Equation~\eqref{eq:clipped-poison}; the perturbation is applied in pixel space, the resulting image is clipped to $[0,1]$, and its target label is retained. For the default Checkerboard attack, the implementation fixes the phase $g_{u,v}=(-1)^{u+v}$ and replicates it over the three RGB channels before clipping. Comparator-specific channel constructions are defined below.

We distinguish three activation settings throughout the paper. In the \emph{native} setting, the test-time magnitude equals the training-time magnitude, $\gamma=1$, and these results provide the primary attack comparison. In the \emph{matched test-time amplification} setting of Section~\ref{sec:attack-performance}, each trained poisoned model and its training magnitude remain fixed while the corresponding test trigger is scaled to $\gamma=1.5$. This standardized additional evaluation is applied to every method with an explicitly scalable trigger; BAAT is not included because its semantic edit has no norm-bounded trigger magnitude. The from-scratch architecture, pretrained fine-tuning, and poison-count/trigger-strength studies additionally vary $\gamma$ for Checkerboard over the values stated in their tables and figures.

\subsection{Comparator Pattern Construction}

The matched study in Section~\ref{sec:pattern-study} is conducted on CIFAR-10. Random-noise entries are drawn from the continuous uniform distribution on $[-1,1]$ independently in RGB space. Salt-and-pepper entries are drawn from $\{-1,+1\}$ in the spatial domain and replicated across RGB. The period-2 stripe patterns are $g^{\rm H}_{u,v}=(-1)^u$ and $g^{\rm V}_{u,v}=(-1)^v$, the block pattern is $g^{\rm B}_{u,v}=(-1)^{\lfloor u/2\rfloor+\lfloor v/2\rfloor}$, and the pixel checkerboard is $g^{\rm C}_{u,v}=(-1)^{u+v}$. Appendix~\ref{app:pattern-constructions} derives the empirical, stripe, and block constructions.

For both empirical patterns, all 50,000 clean CIFAR-10 training images are represented as raw, unaugmented vectors in $[0,1]^{3072}$, centered by their empirical mean, and used without labels. No test image enters the estimate. We compute the ordinary sample covariance $S$ in Equation~\eqref{eq:empirical-rgb-precision} once and use the same raw-sample precision $S^{-1}$ for both constructions. Equation~\eqref{eq:empirical-eq5} operates directly in RGB space; Equation~\eqref{eq:empirical-eq6} uses the induced matrix $R^\top S^{-1}R$ and does not estimate a separate grayscale covariance.

For each empirical objective, we orient the leading eigenvector so that its coordinate of largest absolute magnitude is positive, with ties resolved by flattened coordinate order; we map zero coordinates to $+1$ and sign-round the remaining coordinates. Starting from this binary vector, we repeatedly flip the coordinate with the largest strictly positive objective gain, breaking gain ties by flattened coordinate order, until no improving one-bit flip remains. This procedure uses only the empirical quadratic objective, not victim-model accuracy or ASR. The resulting patterns are reproducible one-flip local optima rather than certified global solutions of the box-constrained problems.

\subsection{Victim Training and Preprocessing}

All evaluation runs use a single NVIDIA RTX~4090. The default victim is ResNet-18~\cite{he2016deep}. From-scratch models are trained for 100 epochs with SGD, momentum 0.9, batch size 256, and cosine decay from an initial learning rate of 0.1. Weight decay is $5\times10^{-4}$ on CIFAR-10, CIFAR-100, and CelebA-8 and $1\times10^{-4}$ on IN-100. All methods compared in the same setting use the same victim architecture, optimization schedule, preprocessing pipeline, target class, and poison count unless explicitly stated otherwise.

We use dataset-specific preprocessing at the resolutions reported in Table~\ref{tab:dataset-budget}: $32\times32$ for CIFAR-10 and CIFAR-100, $64\times64$ for CelebA-8, and $224\times224$ for IN-100. Standard training-time spatial augmentation and normalization are applied consistently to clean, poisoned, and baseline inputs. Poisoning is performed and clipped in pixel space before this augmentation and normalization pipeline.

\subsection{Pretrained Fine-Tuning}

The pretrained-model study evaluates ImageNet-1k-pretrained ResNet-50, ConvNeXt-S, and Swin-T on IN-100. For all three backbones, the poisoning configuration uses target 0, uniform random target-class poison selection, $m=520$, and $\alpha=16/255$. We load ImageNet-1k-pretrained checkpoints, replaces the classifier with a randomly initialized 100-way head, and fine-tunes all parameters using AdamW with learning rate $10^{-4}$, weight decay $0.05$, batch size 64, and 100 epochs. Each reported result is averaged over five independent poisoned-model fine-tuning runs. Native evaluation uses $\alpha=16/255$; the same fine-tuned models are also evaluated at $\{20,24\}/255$ test-time trigger perturbation strength for trigger amplification analysis. The IN-100 classes were seen during full-ImageNet pretraining, so this experiment measures continued supervised fine-tuning rather than transfer to unseen categories.

\subsection{Targets, Runs, Metrics, and Aggregation}

Unless otherwise specified, experiments use target class 0. The multi-target study instead uses all ten CIFAR-10 targets, all eight CelebA-8 targets, and targets 0--9 for both CIFAR-100 and IN-100. Reported cells are means over five independent runs unless explicitly stated otherwise.

ACC is computed on the complete clean test set. To compute ASR, we exclude all test inputs whose ground-truth label is the target class, apply the trigger to the remaining inputs, compute the fraction classified as the target separately for each non-target source class, and take the mean of these source-class rates. In the multi-target experiments, we first average the five runs within each target and then take the mean across the evaluated targets.

\subsection{Poison-Generation Timing}

Poison-generation runtime is measured as wall-clock time on one NVIDIA RTX 4090 and reported as the mean over five runs. The measured interval includes trigger construction, poison creation, and any attack-specific surrogate training or iterative optimization required to generate the poisoned training set. It excludes victim-model training. For BAAT, it also excludes generative-model pretraining, so the reported BAAT value understates its end-to-end setup cost and is not directly comparable to methods whose full attack-specific computation lies inside the measured interval.

\section{Full Multi-Target Results}
\label{app:multi-target}

Table~\ref{tab:targets-all} reports every target used in Table~\ref{tab:multi-target}. CIFAR-10 uses all ten classes, CelebA-8 uses all eight classes, and CIFAR-100 and IN-100 use targets 0--9. Each ACC and ASR entry is averaged over five seeded runs. The main-paper summary takes the equal-weight mean of these per-target values.

\begin{table}[H]
	\centering
	\caption{Per-target Checkerboard results under the dataset-specific configurations in Table~\ref{tab:dataset-budget}. CIFAR-10, CIFAR-100, and CelebA-8 use $\alpha=10/255$ and $\gamma=1$, while IN-100 uses $\alpha=16/255$ and $\gamma=1$. ACC and ASR are percentages averaged over five seeded runs. A dash indicates that CelebA-8 has no corresponding target.}
	\label{tab:targets-all}
	\small
	\setlength{\tabcolsep}{3pt}
	\resizebox{\columnwidth}{!}{%
		\begin{tabular}{@{}ccc@{\hspace{1em}}ccc@{\hspace{1em}}ccc@{\hspace{1em}}ccc@{}}
			\toprule
			\multicolumn{3}{c}{CIFAR-10} &
			\multicolumn{3}{c}{CIFAR-100} &
			\multicolumn{3}{c}{CelebA-8} &
			\multicolumn{3}{c}{IN-100} \\
			\cmidrule(lr){1-3}
			\cmidrule(lr){4-6}
			\cmidrule(lr){7-9}
			\cmidrule(lr){10-12}
			$t$ & ACC & ASR &
			$t$ & ACC & ASR &
			$t$ & ACC & ASR &
			$t$ & ACC & ASR \\
			\midrule
			0 & 94.55 & 95.72 & 0 & 76.66 & 96.39 & 0 & 79.24 & 100 & 0 & 69.94 & 83.76 \\
			1 & 94.53 & 90.35 & 1 & 76.61 & 97.89 & 1 & 79.21 & 100 & 1 & 70.12 & 79.68 \\
			2 & 94.61 & 94.56 & 2 & 76.54 & 98.73 & 2 & 78.99 & 100 & 2 & 70.23 & 81.53 \\
			3 & 94.26 & 96.17 & 3 & 76.68 & 95.91 & 3 & 79.35 & 100 & 3 & 69.97 & 80.37 \\
			4 & 94.28 & 92.13 & 4 & 76.45 & 96.87 & 4 & 79.42 & 100 & 4 & 68.98 & 82.95 \\
			5 & 94.24 & 91.05 & 5 & 76.59 & 98.56 & 5 & 79.18 & 100 & 5 & 69.52 & 84.61 \\
			6 & 94.54 & 90.59 & 6 & 76.62 & 99.98 & 6 & 79.21 & 100 & 6 & 69.61 & 85.82 \\
			7 & 94.56 & 91.58 & 7 & 76.69 & 97.62 & 7 & 79.56 & 100 & 7 & 70.19 & 83.57 \\
			8 & 94.51 & 91.44 & 8 & 76.60 & 95.65 & -- & -- & -- & 8 & 70.05 & 87.63 \\
			9 & 94.24 & 96.78 & 9 & 76.55 & 98.36 & -- & -- & -- & 9 & 69.84 & 79.25 \\
			\bottomrule
		\end{tabular}%
	}
\end{table}

\section{CIFAR-10 Published-Setting Comparison at 500 Poisons}

\label{app:cifar10-fixed500}

We report each baseline under its published applicable CIFAR-10 setting and fix the number of poisoned training samples at $m=500$. Checkerboard is not retuned to the perturbation model or magnitude of each row: it uses the same reference configuration throughout, with $m=500$ and $\alpha=16/255$. The table therefore provides a fixed-poison-count, published-setting reference comparison rather than a perturbation-budget-matched ranking.

\begin{table}[H]
	\centering
	\caption{CIFAR-10 ASR with 500 poisoned training samples
		($m/N=1\%$ and $m/N_t=10\%$). Baselines use their published
		applicable settings; Checkerboard uses $\alpha=16/255$.}
	\label{tab:original-setting-baselines}
	\small
	\renewcommand{\arraystretch}{1.08}
	
	\begin{tabular*}{\columnwidth}{@{\extracolsep{\fill}}lr@{}}
		\toprule
		\textbf{Method} & \textbf{ASR} \\
		\midrule
		LC               & 23.12\% \\
		SIG              & 92.5\% \\
		Invisible Poison & 73.1\% \\
		Narcissus        & 50.1\% \\
		GenBound         & 93.1\% \\
		Hybrid           & 90.53\% \\
		COMBAT           & 92.18\% \\
		SAA              & 61.3\% \\
		BAAT             & 95.12\% \\
		\midrule
		\textbf{Checkerboard} & \textbf{100\%} \\
		\bottomrule
	\end{tabular*}
\end{table}

Table \ref{tab:original-setting-baselines} summarizes the result. Among the baselines, BAAT and GenBound have the highest ASRs at $95.12\%$ and $93.1\%$, followed closely by SIG at $92.5\%$. The fixed Checkerboard reference reaches $100\%$, exceeding the completed baseline values by $4.88\%$--$76.88\%$. 

SAA is source-specific in its standard formulation, and its reported ASR is computed only on triggered inputs from the designated source class; it is not a universal all-to-one setting. BAAT instead edits semantic attributes and is not governed by an $\ell_\infty$ perturbation budget. These two rows broaden methodological coverage but should not be interpreted as scope- or perturbation-matched comparisons with Checkerboard.

\section{Additional Defenses} \label{app:additional-defenses}
\subsection{Model-level Defenses}
\textbf{Neural Cleanse} (SP 19) \cite{wang2019neural} reverse-engineers a class-specific trigger and flags a model as backdoored when one class yields an anomalously small recovered trigger. Following the original protocol, we evaluate the anomaly index of the Checkerboard-poisoned model on both CIFAR-10 and IN-100 and obtain a value of 0.96 and 1.12, respectively, which is well below the standard detection threshold of 2. Therefore, Neural Cleanse does not flag the model. This is because Neural Cleanse is biased toward sparse, localized triggers through its $L_1$-regularized objective, whereas Checkerboard uses a dense, low-amplitude $L_\infty$ perturbation across the entire image. This mismatch makes the recovered pattern unlikely to appear as an outlier.

\textbf{DeBackdoor} (USENIX Security 25) \cite{10.5555/3766078.3766408} is a black-box detector that synthesizes triggers through simulated annealing and gradient-free search over predefined trigger templates. A model is flagged as infected when the synthesized trigger for any class achieves a continuous attack success rate (cASR) above a high threshold, typically 0.90. We evaluate DeBackdoor in its \texttt{lira} \cite{doan2021lira} mode, which assumes an additive invisible trigger consistent with our threat model. For the true target class on CIFAR-10 and IN-100, the synthesized trigger achieves only 7\% and 0.6\% cASR, respectively, far below the detection threshold, and the model is therefore declared backdoor-free. This suggests that the defense’s gradient-free search procedure is unable to reconstruct the global high-frequency structure required by Checkerboard.

\begin{figure}[t]
	\centering
	\includegraphics[width=\linewidth]{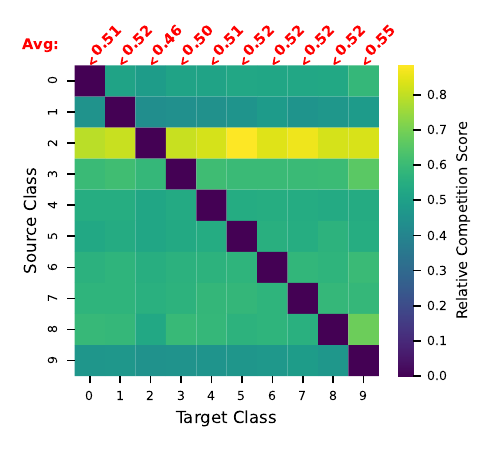}
	\caption{RCS matrix of BARBIE on CIFAR-10.}
	\label{fig:barbie_failure}
\end{figure}

\subsubsection{Model Detection-based Defenses}
\textbf{BARBIE} (NDSS 25) \cite{zhang2025barbie} detects backdoored models by analyzing the dominance of inverted latent features across classes. It computes a Relative Competition Score (RCS) matrix and flags a model when one class exhibits an anomalously dominant score. Due to the heavy computation cost, we evaluate BARBIE on CIFAR-10 Checkerboard-infected models only. We evaluate BARBIE using its official implementation on 200 benign models and 20 Checkerboard-poisoned models on CIFAR-10. Across five runs, BARBIE achieves a true positive rate of only 2\% and a false positive rate of 7.3\%, indicating ineffective detection. The RCS matrix in Figure~\ref{fig:barbie_failure} shows that the target class does not stand out as a clear outlier; its mean score is approximately 0.51, which lies within the benign range. This suggests that under the clean-label and low poisoning budget of Checkerboard, the target class does not become dominant in the inverted feature space for BARBIE to detect it.

\begin{figure}[t]
	\centering
	
	\includegraphics[width=\linewidth]{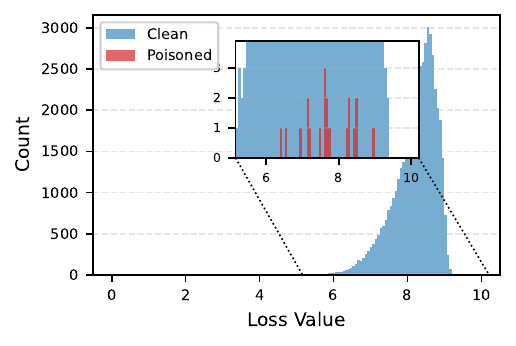}
	\caption{Loss values of ESTI on CIFAR-10}
	
	\label{fig:ESTI}
\end{figure}

\textbf{ESTI} (ICCV 25) \cite{Yu_2025_ICCV} alternates between dual-model training and dynamic loss thresholding to split the dataset into clean and poisoned subsets. It then isolates suspected poisons into a trap class so that triggered inputs are mapped to a harmless label at test time. Following the official implementation and using 1\% clean samples for initialization, ESTI achieves 94.61\% ACC while retaining a 97.23\% ASR against Checkerboard on CIFAR-10. On IN-100, it achieves 68.91\% ACC with an ASR of 82.56\% at $\gamma=1$. Hence, the defense preserves clean performance but does not mitigate the attack. Figure~\ref{fig:ESTI} shows that the poisoned samples do not form a distinct low-loss mode, which is the separation signal ESTI relies on. Under Checkerboard’s setting, poisoned samples remain statistically similar to clean ones in loss space, preventing reliable splitting.

\textbf{ANP} (NeurIPS 21) \cite{wu2021adversarial} identifies and prunes suspect neurons by measuring how sensitively they react to adversarial parameter perturbations. Applied to Checkerboard-poisoned models, ANP preserves clean accuracy at 94.15\% on CIFAR-10 but leaves the ASR nearly unchanged at 95.51\%. On IN-100, it achieves 68.87\% ACC while retaining a high ASR of 83.68\%. Thus, it fails to remove the backdoor. This suggests that the backdoor behavior induced by Checkerboard is not concentrated in a small set of highly perturbation-sensitive neurons, which is the assumption ANP is designed to exploit.

\textbf{D3} (CVPR 25) \cite{wei2025backdoor} aims to remove backdoors by pushing model parameters away from their pre-trained state, under the assumption that backdoor behavior resides in a sharp local minimum. On Checkerboard-poisoned models, D3 achieves 93.85\% ACC while retaining a 95.61\% ASR on CIFAR-10. On IN-100, it achieves 68.45\% ACC with an ASR of 84.15\% at $\gamma=1$. This suggests that Checkerboard does not fall into the sharp-minimum regime that D3 is designed to counter.

\subsection{Data-oriented Defenses}
\subsubsection{Latent Representation Analysis}
\textbf{Spectral Signatures} (NeurIPS 18) \cite{tran2018spectral} identifies poisoned samples by performing singular value decomposition on class-wise deep representations and removing samples with the largest outlier scores along the top spectral direction. The defense implicitly assumes that the poisoned subset forms a spectrally separable sub-population whose mean shift is sufficiently strong relative to within-class variance. However, when evaluated against Checkerboard, Spectral Signatures achieves a 0\% TPR with a 15\% FPR on CIFAR-10, and a 0\% TPR with a 19\% FPR on IN-100, indicating that the underlying separability assumption does not hold for Checkerboard. Specifically, Checkerboard's low-intensity clean-label perturbation does not induce a sufficiently distinct poisoned feature subpopulation, so the resulting features lack a statistically salient spectral signature. Instead, they remain broadly overlapped with the clean target-class distribution. Consequently, the leading singular direction is governed mainly by benign semantic variation rather than by the poisoning signal, so the method removes clean samples aligned with natural class variation while failing to isolate the actual poisoned ones.

\begin{figure}[t]
	
	\centering
	\includegraphics[width=\linewidth]{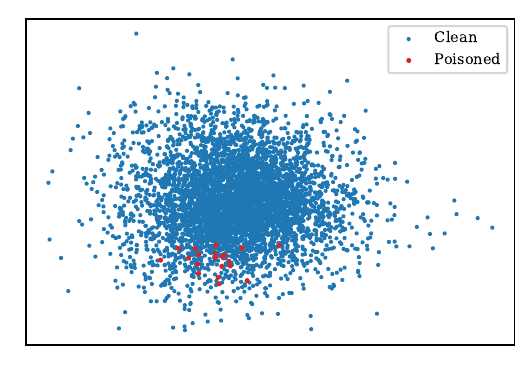}
	
	\caption{Deep feature visualization of SCAn on CIFAR-10.}
	\label{fig:scan}
\end{figure}

\textbf{SCAn} (USENIX Security 21) \cite{tang2021demon} models deep representations through a decomposition into a class-specific identity component and a class-independent variation component, and detects an infected class when its features are better explained by a two-subgroup mixture than by a single class distribution. Following the original setting, we use 1\% clean samples to estimate the decomposition parameters, and the deep features are extracted from the AveragePooling layer of ResNet-18. On the Checkerboard-poisoned model, SCAn yields anomaly scores of 0.48 on CIFAR-10 and 0.31 on IN-100, which are far below the recommended threshold of $e^2 \approx 7.39$, and therefore fails to identify the true target class. Figure~\ref{fig:scan} shows the target-class deep representations projected onto the subspace spanned by the first two principal components. We can observe that the poisoned samples do not form the distinct subgroup structure that SCAn expects. Instead, they remain heavily overlapped with the clean target-class distribution, with only a slight local shift.

\textbf{FLARE} (TIFS 25) \cite{hou2025flare} is a multi-layer defense by aggregating extreme activations from Batch Normalization-aligned feature maps across the entire network to construct comprehensive latent representations. It identifies poisoned samples by projecting these high-dimensional features into a subspace via UMAP \cite{McInnes2018} and using HDBSCAN \cite{McInnes2017} to detect clusters that deviate from benign distributions. We evaluated FLARE against the Checkerboard following the paper's default settings and observed a 0\% TPR on both CIFAR-10 and IN-100. This is because HDBSCAN's density-based clustering is optimized for cohesive outliers, whereas the global, low-intensity Checkerboard trigger produces semantically fragmented feature anomalies that are pulled between the diverse original semantic features and the weak trigger signal.  As a result, these scattered poisoned samples fail to form a sufficiently dense cluster and are classified as noise, which FLARE’s decision rule treats as benign to minimize false positives, making the poisoned samples to remain undetected.

\noindent \textbf{CT} (USENIX Security 22) \cite{Qi2022TowardsAP} is a proactive backdoor defense framework that intentionally decouples benign data-label correlations by training jointly with randomly mislabeled clean samples. It forces the resulting inference model to be difficult in fitting normal clean data (yielding high loss) while retaining its ability to fit backdoor poison samples (yielding low loss), enabling their separation through an iterative loss-based distillation. Specifically, CT first attempts to identify suspicious target classes by applying a Gaussian Mixture Model clustering analysis to the latent space; only classes exceeding a predefined likelihood ratio threshold (typically 2.0) are subjected to the final sample-level detection. We evaluated CT against Checkerboard attack with 4\% clean samples as mentioned in its paper. We find that CT cannot correctly identify the target class and yield a TPR of 0.0\% and a FPR of 1.6\% on CIFAR-10, and a TPR of 3.5\% and 4.3\% FPR on IN-100. CT cannot detect the Checkerboard injected samples because its trigger is constrained by a low infinity norm relatively to the backgrounds of each dataset. During the confusion training, this weak artifact is easily drowned out by the artificially induced label noise; the model cannot learn the backdoor correlation and consequently assigns high losses to the poison samples. As a result, the iterative distillation process falsely treats the poison samples as unfittable clean data instead.

\begin{figure}[t]
	\centering
	\includegraphics[width=\linewidth]{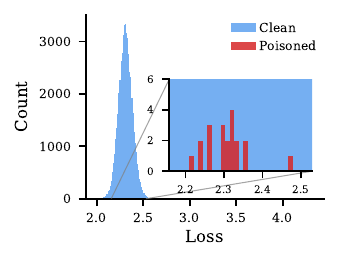}
	\caption{ASSET's loss values on CIFAR-10.}
	\label{fig:ASSET}
\end{figure}

\textbf{ASSET} (USENIX Security 23) \cite{pan2023asset} uses a nested min-max optimization procedure to amplify behavioral differences between clean and poisoned samples, with the goal of pushing poisoned samples into a high-loss region that can be separated from clean data. Following the original setting, we use 1,000 clean samples to guide the optimization. On Checkerboard, ASSET achieves only 25.0\% TPR while incurring 38.53\% FPR on CIFAR-10, and 11.21\% TPR and 25.67\% FPR on IN-100. Thus, it fails to isolate poisoned samples without causing substantial collateral damage. Figure~\ref{fig:ASSET} shows that the loss values of clean and poisoned samples remain highly overlapping, which is consistent with the strong clean/poison loss overlap observed for Checkerboard.

\begin{figure}[t]
	\includegraphics[width=\linewidth]{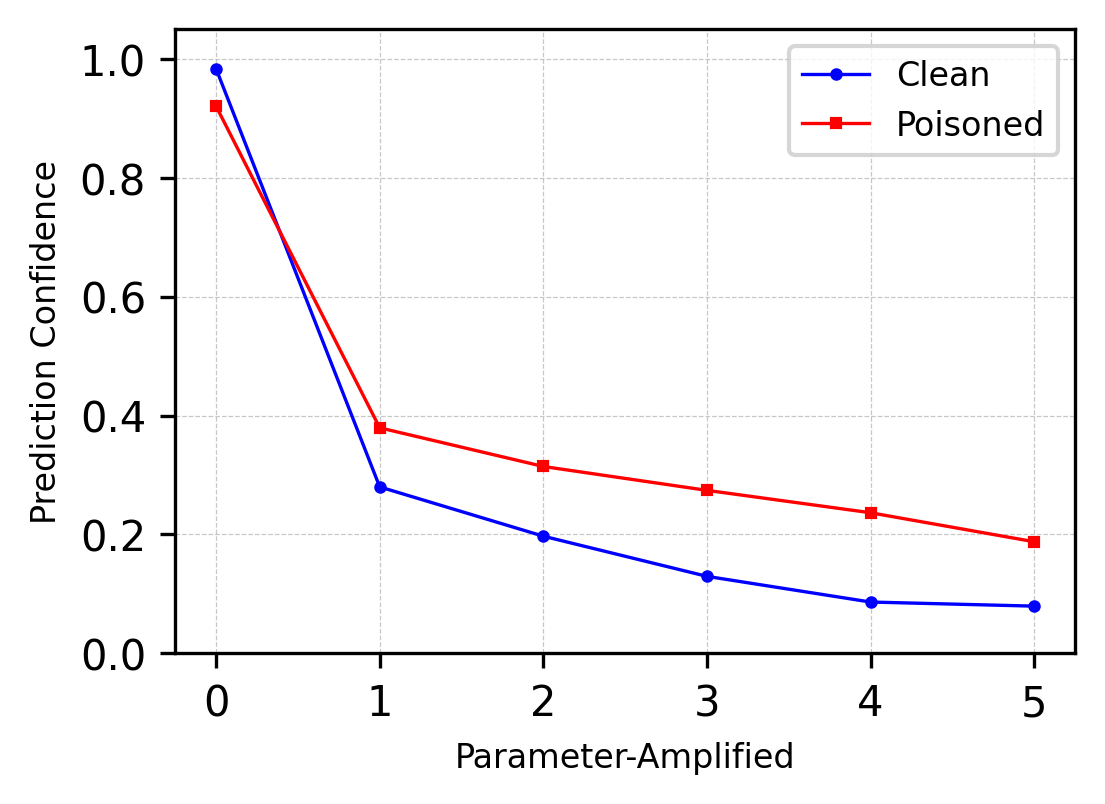}
	\caption{PSC curve of IBD-PSC on CIFAR-10.}
	\label{fig:IBD-PSC}
	
\end{figure}

\textbf{IBD-PSC} (ICML 24) \cite{10.5555/3692070.3692834} is an state-of-the-art black-box test-time detector that measures parameter-oriented scaling consistency (PSC). It scales selected BatchNorm layers, tracks the confidence of the original prediction across the perturbed models, and flags an input when the average confidence exceeds a threshold. Following the original setting with 0.2\% clean samples, we evaluate IBD-PSC on Checkerboard and obtain TPR is only 2.7\% and FPR is 5.15\% on CIFAR-10, and 1.5\% TPR and 1.98\% FPR on IN-100. Figure \ref{fig:IBD-PSC} illustrates this failure by tracking prediction confidences across 5 BN amplification steps. Contrary to IBD-PSC's core assumption, the confidence of Checkerboard-poisoned samples behaves similarly to that of benign samples, which causes this defense fail.

\subsection{Adaptive Defenses}
\label{sec_appendix_adaptive_defense}

In this section, we provide additional details and results for the adaptive defenses discussed in Section~\ref{sec:adaptive-defenses}. These defenses are designed with knowledge of Checkerboard's mechanisms and therefore represent stronger evaluations than generic defense settings. We consider two types of adaptive defenses: pattern-aware input sanitization and preprocessing-based high-frequency suppression..

\subsubsection{Checkerboard-Notch}
\label{sec_appendix_checkerboard_notch}

Checkerboard-Notch explicitly targets the known spatial mode of the checkerboard trigger. Let \(\mathbf{x}\in[0,1]^{H\times W\times C}\) denote an input image, where \(H\times W\) is the spatial resolution and \(C\) is the number of channels. We first define the spatial checkerboard basis
\[
q_{i,j}=(-1)^{i+j}, \qquad 1\leq i\leq H,\; 1\leq j\leq W.
\]
For color images, this spatial pattern is replicated across all channels to form the channel-expanded checkerboard template \(\mathbf{Q}\in\mathbb{R}^{H\times W\times C}\). This template captures the alternating high-frequency component used by the Checkerboard trigger.

For each input image, we estimate the checkerboard coefficient by projecting the image onto the checkerboard template:
\[
c(\mathbf{x})=\frac{\langle \mathbf{x},\mathbf{Q}\rangle}{\|\mathbf{Q}\|_2^2},
\]
where \(\langle\cdot,\cdot\rangle\) denotes the inner product over all spatial locations and channels. Intuitively, \(c(\mathbf{x})\) measures the strength of the checkerboard component contained in the input.

To avoid unnecessarily modifying samples with only weak checkerboard correlation, we apply a soft-thresholding rule:
\[
\hat{c}(\mathbf{x})=
\begin{cases}
	0, & |c(\mathbf{x})|\leq \tau,\\[4pt]
	\operatorname{sign}(c(\mathbf{x}))\big(|c(\mathbf{x})|-\tau\big), & |c(\mathbf{x})|>\tau,
\end{cases}
\]
where \(\tau\geq0\) is a threshold parameter. The sanitized image is then obtained by subtracting the estimated checkerboard component:
\[
\tilde{\mathbf{x}}
=
\Pi_{[0,1]}
\Big(
\mathbf{x}-\lambda \hat{c}(\mathbf{x})\mathbf{Q}
\Big),
\]
where \(\lambda>0\) controls the suppression strength and \(\Pi_{[0,1]}(\cdot)\) denotes element-wise clipping to the valid pixel range. When \(\lambda=1\), the estimated checkerboard component is fully removed; smaller values of \(\lambda\) yield more conservative suppression.

\begin{figure}[t]
	\centering
	\begin{subfigure}{0.32\columnwidth}
		\centering
		\includegraphics[width=\linewidth]{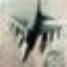}
	\end{subfigure}
	\begin{subfigure}{0.32\columnwidth}
		\centering
		\includegraphics[width=\linewidth]{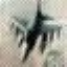}
	\end{subfigure}
	\begin{subfigure}{0.32\columnwidth}
		\centering
		\includegraphics[width=\linewidth]{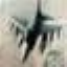}
	\end{subfigure}
	
	\caption{Comparison of samples under the Checkerboard-Notch adaptive defense on CIFAR-10. From left to right: clean sample, poisoned sample, and poisoned sample sanitized by Checkerboard-Notch.}
	\label{fig:checkerboard-notch}
\end{figure}

In our reported adaptive-defense evaluation, we use the strongest setting with \(\tau=0\) and \(\lambda=1\), which removes the full estimated checkerboard projection from each input. An example sanitized poisoned sample is shown in Figure~\ref{fig:checkerboard-notch}. As illustrated, Checkerboard-Notch removes most of the visually observable checkerboard component. However, as reported in Section~\ref{sec:adaptive-defenses}, this pattern-aware sanitization only weakens the default test-time trigger and does not eliminate the learned backdoor behavior: the ASR recovers strongly under moderate test-time trigger amplification. This indicates that directly subtracting the estimated checkerboard component is insufficient to fully erase the model's reliance on the checkerboard-induced signal.

\subsubsection{Additional Preprocessing}
\label{sec_appendix_preprocessing}

\begin{table}[t]
	\centering
	\caption{Gaussian-blur adaptive-defense results.}
	\label{tab:adaptive-defense}
	\small
	\setlength{\tabcolsep}{4.0pt}
	\begin{tabular}{rrrcrrrr}
		\toprule
		\multicolumn{8}{l}{\textbf{CIFAR-10}}\\
		$b$ & $m$ & Train $\alpha$ & $\sigma$ & ACC & \multicolumn{3}{c}{ASR} \\
		\cmidrule(lr){6-8}
		& & & & & $10/255$ & $15/255$ & $20/255$ \\
		\midrule
		\multirow[c]{3}{*}{1} & \multirow[c]{3}{*}{20} & \multirow[c]{3}{*}{$10/255$}
		& 0.5 & 94.29 & 95.97 & 99.97 & 100.00 \\
		& & & 1.0 & 93.08 & 0.67 & 0.67 & 0.67 \\
		& & & 2.0 & 92.47 & 0.85 & 0.94 & 1.04 \\
		\addlinespace[1pt]
		\multirow[c]{3}{*}{2} & \multirow[c]{3}{*}{100} & \multirow[c]{3}{*}{$10/255$}
		& 0.5 & 94.57 & 43.50 & 79.21 & 90.91 \\
		& & & 1.0 & 93.11 & 70.84 & 97.51 & 99.84 \\
		& & & 2.0 & 93.08 & 51.31 & 89.00 & 98.42 \\
		\midrule
		\multicolumn{8}{l}{\textbf{IN-100}}\\
		$b$ & $m$ & Train $\alpha$ & $\sigma$ & ACC & \multicolumn{3}{c}{ASR} \\
		\cmidrule(lr){6-8}
		& & & & & $16/255$ & $20/255$ & $24/255$ \\
		\midrule
		\multirow[c]{3}{*}{1} & \multirow[c]{3}{*}{520} & \multirow[c]{3}{*}{$16/255$}
		& 0.5 & 69.5 & 80.26 & 90.81 & 96.58 \\
		& & & 1.0 & 67.68 & 12.53 & 30.51 & 42.31 \\
		& & & 2.0 & 67.13 & 9.68 & 23.15 & 33.16 \\
		\addlinespace[1pt]
		\multirow[c]{3}{*}{2} & \multirow[c]{3}{*}{650} & \multirow[c]{3}{*}{$16/255$}
		& 0.5 & 69.52 & 60.51 & 75.56 & 89.65 \\
		& & & 1.0 & 67.71 & 55.63 & 69.81 & 84.39 \\
		& & & 2.0 & 67.05 & 53.25 & 67.44 & 83.82 \\
		\bottomrule
	\end{tabular}
\end{table}

We further evaluate preprocessing-based adaptive defenses that suppress high-frequency image content. These defenses include mean filtering, JPEG compression, and DCT-based high-frequency suppression. Throughout this subsection, we use the checkerboard block size \(b\) to describe the spatial granularity of the alternating pattern. \(b=1\) denotes the default pixel-wise checkerboard, where adjacent pixels alternate between \(+1\) and \(-1\). \(b=2\) denotes a lower-frequency variant $2 \times 2$ block checkerboard trigger pattern, where the sign alternates every two pixels in both spatial dimensions.

Consistent with the main-text adaptive evaluation, the \(b=1\) setting uses the default Checkerboard configuration, while the \(b=2\) setting corresponds to the adaptive lower-frequency variant. The lower-frequency variant reduces sensitivity to low-pass preprocessing. Since it sacrifices part of the input-space separability advantage of the default pixel-wise checkerboard, it uses a moderately larger, but still low, poison-sample budget while keeping the perturbation magnitude fixed.


\paragraph{Mean Filtering.}
Mean filtering replaces each pixel with the average of its local neighborhood, thereby suppressing local high-frequency variations. We evaluate \(3\times3\) and \(5\times5\) mean filters, with results reported in Table~\ref{tab:mean_filter}. On CIFAR-10, the default \(b=1\) trigger remains effective under the \(3\times3\) filter, achieving 65.07\% ASR at \(\gamma=1\) and 88.57\% when \(\gamma\alpha=20/255\). The stronger \(5\times5\) filter suppresses this configuration, reducing its ASR to approximately 1\%. However, the adaptive \(b=2\) variant recovers 72.64\% ASR under the \(5\times5\) filter when \(\gamma\alpha=20/255\). A similar pattern appears on IN-100. Under the \(5\times5\) filter, the default trigger achieves only 10.53\% ASR at \(\gamma=1\) and 18.32\% when \(\gamma\alpha=24/255\), whereas the \(b=2\) variant reaches 43.64\% and 75.61\%, respectively. Under the weaker \(3\times3\) filter, both variants remain effective at the largest evaluated test-time magnitude, reaching 83.69\% and 86.65\% ASR. These results show that strong mean filtering can suppress the default pixel-wise trigger, while reducing the trigger’s spatial frequency substantially improves robustness under aggressive smoothing.

\begin{table}[t]
	\centering
	\caption{Mean-filtering adaptive-defense results.}
	\label{tab:mean_filter}
	\small
	\setlength{\tabcolsep}{4.0pt}
	\begin{tabular}{rrrcrrrr}
		\toprule
		\multicolumn{8}{l}{\textbf{CIFAR-10}}\\
		$b$ & $m$ & Train $\alpha$ & Kernel & ACC & \multicolumn{3}{c}{ASR} \\
		\cmidrule(lr){6-8}
		& & & & & $10/255$ & $15/255$ & $20/255$ \\
		\midrule
		1 & 20  & $10/255$ & $3\times3$ & 93.00 & 65.07 & 82.89 & 88.57 \\
		1 & 20  & $10/255$ & $5\times5$ & 88.41 & 1.18 & 1.26 & 1.36 \\
		2 & 100 & $10/255$ & $3\times3$ & 93.18 & 23.92 & 56.62 & 81.17 \\
		2 & 100 & $10/255$ & $5\times5$ & 89.80 & 15.30 & 43.97 & 72.64 \\
		\midrule
		\multicolumn{8}{l}{\textbf{IN-100}}\\
		$b$ & $m$ & Train $\alpha$ & Kernel & ACC & \multicolumn{3}{c}{ASR} \\
		\cmidrule(lr){6-8}
		& & & & & $16/255$ & $20/255$ & $24/255$ \\
		\midrule
		1 & 520 & $16/255$ & $3\times3$ & 67.18 & 50.23 & 75.82 & 83.69 \\
		1 & 520 & $16/255$ & $5\times5$ & 67.53 & 10.53 & 15.66 & 18.32 \\
		2 & 650 & $16/255$ & $3\times3$ & 66.16 & 55.61 & 78.89 & 86.65 \\
		2 & 650 & $16/255$ & $5\times5$ & 66.04 & 43.64 & 60.25 & 75.61 \\
		\bottomrule
	\end{tabular}
\end{table}

\begin{table}[t]
	\centering
	\caption{JPEG-compression adaptive-defense results.}
	\label{tab:jpeg}
	\small
	\setlength{\tabcolsep}{4.0pt}
	\begin{tabular}{rrrcrrrr}
		\toprule
		\multicolumn{8}{l}{\textbf{CIFAR-10}}\\
		$b$ & $m$ & Train $\alpha$ & Quality & ACC & \multicolumn{3}{c}{ASR} \\
		\cmidrule(lr){6-8}
		& & & & & $10/255$ & $15/255$ & $20/255$ \\
		\midrule
		1 & 20  & $10/255$ & 50 & 89.73 & 0.85 & 0.74 & 0.95 \\
		1 & 20  & $10/255$ & 75 & 91.49 & 0.80 & 0.82 & 0.78 \\
		1 & 20  & $10/255$ & 85 & 92.41 & 0.64 & 0.63 & 0.64 \\
		1 & 20  & $10/255$ & 95 & 93.86 & 44.77 & 86.57 & 89.17 \\
		2 & 100 & $10/255$ & 50 & 89.65 & 3.10 & 29.46 & 52.67 \\
		2 & 100 & $10/255$ & 75 & 91.60 & 23.87 & 74.11 & 94.64 \\
		2 & 100 & $10/255$ & 85 & 92.40 & 47.54 & 89.62 & 98.44 \\
		2 & 100 & $10/255$ & 95 & 94.08 & 52.81 & 91.45 & 98.01 \\
		\midrule
		\multicolumn{8}{l}{\textbf{IN-100}}\\
		$b$ & $m$ & Train $\alpha$ & Quality & ACC & \multicolumn{3}{c}{ASR} \\
		\cmidrule(lr){6-8}
		& & & & & $16/255$ & $20/255$ & $24/255$ \\
		\midrule
		1 & 520 & $16/255$ & 50 & 68.00 & 0.12 & 0.11 & 0.11 \\
		1 & 520 & $16/255$ & 75 & 69.42 & 0.21 & 0.19 & 0.20 \\
		1 & 520 & $16/255$ & 85 & 69.53 & 0.16 & 0.16 & 0.15 \\
		1 & 520 & $16/255$ & 95 & 69.26 & 53.68 & 67.83 & 75.96 \\
		2 & 650 & $16/255$ & 50 & 67.98 & 20.66 & 45.37 & 67.86 \\
		2 & 650 & $16/255$ & 75 & 69.51 & 31.61 & 55.68 & 82.53 \\
		2 & 650 & $16/255$ & 85 & 69.82 & 40.83 & 62.31 & 85.67 \\
		2 & 650 & $16/255$ & 95 & 69.44 & 50.12 & 69.87 & 87.96 \\
		\bottomrule
	\end{tabular}
\end{table}


\paragraph{JPEG Compression.}
JPEG compression suppresses high-frequency image components through lossy quantization. We evaluate quality factors 50, 75, 85, and 95, where lower values indicate stronger compression. The results are reported in Table~\ref{tab:jpeg}. On CIFAR-10, compression at qualities 50--85 almost completely suppresses the default \(b=1\) trigger, keeping its ASR below 1\% across all evaluated test-time magnitudes. At quality 95, however, its ASR increases from 44.77\% at \(\gamma=1\) to 89.17\% when \(\gamma\alpha=20/255\). The adaptive \(b=2\) configuration is considerably more resistant: at the same test-time magnitude, it achieves 94.64\%, 98.44\%, and 98.01\% ASR at qualities 75, 85, and 95, respectively, and retains 52.67\% ASR even under quality 50. The IN-100 results exhibit a similar trend. For \(b=1\), qualities 50--85 reduce the ASR to at most 0.21\%, whereas quality 95 retains 53.68\% ASR at \(\gamma=1\) and 75.96\% when \(\gamma\alpha=24/255\). Under the adaptive \(b=2\) configuration, ASR reaches 67.86\% even at quality 50 and ranges from 82.53\% to 87.96\% at qualities 75--95 under the largest evaluated test-time magnitude. These results indicate that aggressive JPEG compression can suppress high-frequency triggers, but the lower-frequency Checkerboard variant remains effective under less destructive compression settings.

\paragraph{DCT-Based High-Frequency Suppression.}
We further evaluate a frequency-domain preprocessing defense based on the discrete cosine transform (DCT). The defense transforms each image into the frequency domain, zeros out the bottom-right \(k\times k\) region containing high-frequency coefficients, and reconstructs the image using the inverse DCT. We evaluate \(k\in\{2,4,8,12,16\}\), where a larger \(k\) removes a larger high-frequency region. The results are reported in Table~\ref{tab:dct}.

On CIFAR-10, the default \(b=1\) trigger remains effective at \(k=2\), reaching 95.53\% ASR when \(\gamma\alpha=20/255\), but is strongly suppressed for \(k\geq4\), where its ASR remains below 4\%. In contrast, the adaptive \(b=2\) configuration achieves 88.43--97.76\% ASR across all evaluated values of \(k\) at the same test-time magnitude. On IN-100, the ASR of the default trigger decreases from 86.34\% at \(k=2\) to 39.56\% at \(k=4\), 5.60\% at \(k=8\), and below 1\% at \(k\geq12\) when \(\gamma\alpha=24/255\). Under the adaptive \(b=2\) configuration, however, ASR remains between 81.67\% and 87.69\% across all suppression sizes, including 82.01\% at \(k=16\). These results show that removing high-frequency DCT coefficients suppresses the default pixel-wise trigger, whereas the lower-frequency configuration substantially improves robustness against explicit frequency-domain suppression.

%

\begin{table}[t]
	\centering
	\caption{DCT-based high-frequency-suppression results. Parameter $k$ is the side length of the zeroed high-frequency region.}
	\label{tab:dct}
	\small
	\setlength{\tabcolsep}{4.0pt}
	\begin{tabular}{rrrcrrrr}
		\toprule
		\multicolumn{8}{l}{\textbf{CIFAR-10}}\\
		$b$ & $m$ & Train $\alpha$ & $k$ & ACC & \multicolumn{3}{c}{ASR} \\
		\cmidrule(lr){6-8}
		& & & & & $10/255$ & $15/255$ & $20/255$ \\
		\midrule
		1 & 20  & $10/255$ & 2  & 94.70 & 47.64 & 84.32 & 95.53 \\
		1 & 20  & $10/255$ & 4  & 94.72 & 1.18  & 2.38  & 3.84 \\
		1 & 20  & $10/255$ & 8  & 94.90 & 0.55  & 0.50  & 0.46 \\
		1 & 20  & $10/255$ & 12 & 94.57 & 0.67  & 0.71  & 0.74 \\
		1 & 20  & $10/255$ & 16 & 93.99 & 0.69  & 0.69  & 0.70 \\
		2 & 100 & $10/255$ & 2  & 94.60 & 36.98 & 78.13 & 92.94 \\
		2 & 100 & $10/255$ & 4  & 94.46 & 51.54 & 88.81 & 97.76 \\
		2 & 100 & $10/255$ & 8  & 94.44 & 47.91 & 81.57 & 92.61 \\
		2 & 100 & $10/255$ & 12 & 94.22 & 37.46 & 73.26 & 90.93 \\
		2 & 100 & $10/255$ & 16 & 93.30 & 25.31 & 65.64 & 88.43 \\
		\midrule
		\multicolumn{8}{l}{\textbf{IN-100}}\\
		$b$ & $m$ & Train $\alpha$ & $k$ & ACC & \multicolumn{3}{c}{ASR} \\
		\cmidrule(lr){6-8}
		& & & & & $16/255$ & $20/255$ & $24/255$ \\
		\midrule
		1 & 520 & $16/255$ & 2  & 69.15 & 50.25 & 78.66 & 86.34 \\
		1 & 520 & $16/255$ & 4  & 69.58 & 15.58 & 27.65 & 39.56 \\
		1 & 520 & $16/255$ & 8  & 68.85 & 2.33 & 3.87 & 5.60 \\
		1 & 520 & $16/255$ & 12 & 68.12 & 0.51 & 0.60 & 0.58 \\
		1 & 520 & $16/255$ & 16 & 68.05 & 0.42 & 0.53 & 0.49 \\
		2 & 650 & $16/255$ & 2  & 69.21 & 45.58 & 78.64 & 87.69 \\
		2 & 650 & $16/255$ & 4  & 69.02 & 40.38 & 75.36 & 84.57 \\
		2 & 650 & $16/255$ & 8  & 68.76 & 35.91 & 73.67 & 83.51 \\
		2 & 650 & $16/255$ & 12 & 68.32 & 33.83 & 74.68 & 81.67 \\
		2 & 650 & $16/255$ & 16 & 67.86 & 30.66 & 74.96 & 82.01 \\
		\bottomrule
	\end{tabular}
\end{table}

Overall, the preprocessing results show a consistent pattern. Strong high-frequency suppression can weaken the default pixel-wise checkerboard, but the lower-frequency adaptive variant substantially restores attack effectiveness. This suggests that these preprocessing defenses mainly exploit the specific frequency choice of the default trigger rather than eliminating the underlying backdoor vulnerability. Moreover, stronger preprocessing often introduces a utility cost, as reflected by reduced clean accuracy under aggressive filtering or compression. Therefore, Checkerboard remains resilient to preprocessing-based adaptive defenses through a simple low-frequency  adjustment.

\end{document}